\documentclass[11pt]{article}

\usepackage{graphicx}
\usepackage{amssymb}
\usepackage{epstopdf}
\usepackage{geometry}
\usepackage{amsmath}

\newcommand{\irr}{\int_{\mathbb R^2}}
\newcommand{\pd}[2]{\frac{\partial #1}{\partial #2}} 
\newcommand{\pbpv}[1]{\left\{ #1, H^N\right\}} 
\newcommand{\bfx}{\mathbf{x}}
\newcommand{\lex}{\left<}
\newcommand{\rex}{\right>}
\newcommand{\hs}{\hspace{1mm}}
\renewcommand{\d}{\partial}  
\newcommand{\rhotilde}{\tilde{\rho}} 
\newcommand{\zetatilde}{\tilde{\zeta}} 
\newcommand{\psitilde}{\tilde{\psi}} 
\newcommand{\la}{\langle} 
\newcommand{\ra}{\rangle} 
\renewcommand{\L}{\mathcal L}
\renewcommand{\H}{\mathcal H} 

\begin{document}

\title{\bf Reduced models of point vortex systems}
\author{ {\bf Jonathan Maack$^{\dagger}$
 and Bruce Turkington$^{*}$ }   \\  \\
Department of Mathematics and Statistics \\ 
 University of Massachusetts Amherst}

\maketitle 

\begin{abstract}  \noindent 
Nonequilibrium statistical models of point vortex systems are constructed
using an optimal closure method, and  these models are employed to approximate the
relaxation toward equilibrium of systems governed by the two-dimensional Euler equations, as well as 
the quasi-geostrophic equations for either single-layer or two-layer  flows.  
Optimal closure refers to a general method of reduction for Hamiltonian systems, in which
macroscopic states are required to belong to  
a parametric family of distributions on phase space.   In the case of point vortex ensembles, the
macroscopic variables describe the spatially coarse-grained vorticity.   
Dynamical closure in terms of those macrostates is obtained by optimizing over paths
in the parameter space of the reduced model subject to the constraints imposed by conserved quantities.   
This optimization minimizes a cost functional that quantifies the rate of information loss due to 
model reduction, meaning that an optimal path represents  a macroscopic evolution that is most
 compatible with the microscopic dynamics in an information-theoretic sense.         
A near-equilibrium linearization of this method is used to derive dissipative equations 
for the low-order spatial moments of ensembles of point vortices in the plane.   
These severely reduced models describe the late-stage evolution of coherent structures 
in two-dimensional and geostrophic turbulence.   For single-layer dynamics, they approximate
the relaxation of initially distorted structures toward axisymmetric equilibrium states.      
For two-layer dynamics, they predict the rate of energy transfer in baroclinically perturbed structures 
returning to stable barotropic states.    
Comparisons against direct numerical simulations of the fully-resolved many-vortex dynamics
validate the predictive capacity of these reduced models.
\end{abstract}   

\vspace{5mm} 
\noindent
Keywords:   vortex dynamics, nonequilibrium statistical mechanics, quasi-geostrophic equations, coherent structures

\vspace{5mm} 

\small  
\noindent 
* Correspondence: turk@math.umass.edu \\   
$\dagger$ Current address: Complex Systems Simulation and Optimization Group, Computational Science Center, 
National Renewable Energy Lab, Golden, CO 80111, USA. jonathan.maack@nrel.gov     
\normalsize 


\newpage

\section{Introduction}
The equilibrium statistical mechanics of point vortex systems now exists as a mature theory.    
Onsager provided the first insight that equilibrium theory is able to explain the organization
of many-vortex systems into coherent structures \cite{onsager}.  He also indicated that  these systems have novel 
thermodynamic properties, most notably that they allow equilibria with negative absolute temperature 
and that the usual equivalence of ensembles between microcanonical and canonical distributions may 
break down.    In the decades since his prescient work, not only have 
these properties been established theoretically and verified numerically, but it has been understood that, 
due to the long-range nature of the interactions between point vortices, the appropriately-scaled continuum limit 
is an exact mean-field theory \cite{CLMP1,CLMP2,eyink-spohn,kiessling}. 
Consequently, the thermodynamic behavior of equilibrium states is completely
determined by a mean-field equation, whose solutions are  steady two-dimensional flows having
a special functional relation between mean vorticity and streamfunction, which is derived from the 
populations of the vortex strengths (circulations) in the statistical ensemble \cite{turk-leshouches,majda-wang, dibattista-majda}.    

Numerical and experimental studies of freely-decaying two-dimensional turbulence have repeatedly demonstrated
that coherent structures emerge and persist within the vorticity fluctuations and filaments of high Reynolds number 
flow, and that  these vortical structures eventually dominate the large scales of the flow 
\cite{mcwilliams1,mcwilliams2,kraichnan-montgomery,tabeling,boffeta-ecke}.    
These coherent structures have naturally been proposed 
as realizations of the equilibrium statistical states of an underlying point vortex dynamics;     for instance, 
direct numerical simulations in doubly-periodic geometry tend to produce the dipolar end states predicted 
by the mean-field theory \cite{montgomery-etal}.  
More recently, alternative statistical equilibrium theories have been developed from other
discretizations of the continuum equations for an incompressible, inviscid fluid, which respect invariants
that the point vortex idealization violates \cite{miller-weichman-cross,robert-sommeria,turk-CPAM}.     
From the point of view of fluid mechanical outcomes, however, these more intricate theories are similar to 
the point vortex theory, in that they also derive special mean-field equations 
satisfied by their equilibrium states.   These theories, when extended to include quasi-geostrophic dynamics, 
have been shown to realize interesting coherent structures such as the zonal jets and embedded Great Red
Spot on Jupiter \cite{TMHD,bouchet-sommeria,majda-wang}.              

By contrast, there is no comparable theory of the nonequilibrium behavior of vortex ensembles, which could furnish 
models of the unsteady behavior of coarse-grained states in two-dimensional or quasi-geostrophic turbulence.    This
is  not surprising in view of the large conceptual gap between equilibrium and nonequilibrium statistical mechanics themselves.
Whereas statistical equilibrium is conceived in the unified framework of Gibbsian ensembles built from exactly 
conserved quantities, nonequilibrium descriptions depend
upon some selection of relevant, or resolved, variables, which are not conserved but are intended to produce 
a coarse-grained dynamics.   Typically the feasibility of such a selection depends on some separation of scales of motion, 
which may or may not be satisfied in problems of interest.    
One systematic approach is through kinetic theory and its hierarchy of reduced distribution functions.  
In this context modeling becomes a choice of a truncation of the hierarchy and  a concomitant closure 
hypothesis, carrying with it restrictions on its range of validity \cite{balescu}.  
An interesting approach to vortex systems through kinetic theory is developed in \cite{chavanis}.       

In this paper we address the nonequilibrium statistical mechanics of point vortex systems by means of 
a less traditional method of model reduction,  which we call ``optimal closure" \cite{turk-JSP,kleeman}.     
Some previous publications have applied this method to other prototypical problems in statistical fluid dynamics 
\cite{kleeman-turk,TT1, TCT, TT2}.  
Our goal is to apply the optimal closure method to model the relaxation of near-equilibrium vortical structures 
toward equilibrium.   Specifically, we seek to extend the range of the mean-field theory known for equilibrium
states to a comparable nonequilibrium mean-field theory of near-equilibrium relaxation.  

The principal merit of the optimal closure method is that, given a reduction defined by some selected
macroscopic observables,  it extracts the intrinsic dissipative equations for the macrostates 
directly from the underlying microscopic dynamics.     
The accuracy of the reduced dynamics depends, of course, upon the choice of the macroscopic description.  
Since a practical model reduction relegates many degrees of freedom to a statistical description, 
there is no guarantee that the derived macroscopic evolution will agree quantitatively with the
ensemble-averaged behavior of the full-resolved microscopic dynamics.  
The problem addressed in the present paper, therefore, offers a useful test of the optimal closure method, 
since it concerns a severe coarse-graining of point vortex dynamics.   For this reason the paper first develops
the reduced model theoretically, and then proceeds to validate it against direct numerical simulations of ensembles.     
 
 For simplicity, we restrict our analysis to isolated structures, that is, ensembles of many, like-signed, equal circulation, 
 point vortices in the entire plane, whose equilibrium states are single macrovortices centered at the origin.          
 The severe coarse-graining we impose takes the macroscopic variables to be 
 the first- or second-order spatial moments of the vorticity.  These moments, together with the conserved energy 
 and circulation, describe the overall ``shape" of the vortex ensemble.      
 For instance, an initially elliptical structure develops spiral arms that wrap around its center and eventually 
 homogenize into an axisymmetric end state.  Our reduced model of this equilibration uses only
 the second-order spatial moments of the vorticity around the center to follow the evolving structure; heuristically
 the reduced model replaces the intricate configuration of point vortices by an effective elliptical distribution
 of vorticity.            
 
 In the first half of the paper we develop the optimal closure theory of point vortex systems governed by 
 inviscid, incompressible Euler equations in the plane.    In the second half we generalize that theory to quasi-geostrophic dynamics, first for barotropic single-layer flows and then for baroclinic two-layer flows.   
 For these simplified geophysical fluids,  we conduct numerical experiments to test the accuracy of 
 the reduced models.   We especially draw attention to the predicted rates of relaxation of vortical structures
 and the dependence of those rates on the key parameters in the quasi-geostrophic equations.     Indeed,    
 the useful outcome of such severe model reduction and closure is to approximate gross features and functional dependences with much greater computational efficiency than is required by direct numerical studies of the full dynamics.


\section{Statistical Mechanics of Point Vortices}

The streamfunction-vorticity formulation of the incompressible, inviscid, two-dimensional 
Euler equation in the plane is  
\begin{subequations} \label{euler-eqn}
\begin{align}
	\partial_t \zeta + [\zeta, \psi] = 0, \label{vorticity-advection} \\
	-\Delta \psi = \zeta, \label{streamfun-eqn}
\end{align}
\end{subequations}
where
$$
	[A, B] = \pd{A}{x}\pd{B}{y} - \pd{A}{y}\pd{B}{x}.
$$
denotes the Poisson bracket on $\mathbb R^2$.   The scalar conservation law \eqref{vorticity-advection}
for the vorticity, $\zeta(\bfx,t)$, states that the vorticity is constant along all flow trajectories, 
$ \dot{x} =  \d \psi / \d y , \;\; \dot{y} = - \d \psi / \d x $.  This is the distinguishing property of two-dimensional
flow \cite{lamb,marchioro-pulvirenti}.   It implies the
conservation of the total circulation, $\Gamma (\zeta)  = \irr \zeta \hs d\bfx $, 
as well as all higher moments of the vorticity (general enstrophy integrals).
The continuum dynamics \eqref{euler-eqn} has a
(noncanonical) Hamiltonian structure, and its Hamiltonian functional, 
which is conserved, is a quadratic functional of $\zeta$  \cite{turk-leshouches}:   
\begin{equation} \label{energy}
	H(\zeta)   =   \frac{1}{2} \irr \psi \,  \zeta \hs d \bfx \, .   
\end{equation}
The translational and rotational symmetries of the domain imply two additional invariants: 
the first and (radial) second spatial moments of vorticity, 
\begin{equation}    
	\mathbf B(\zeta)   = \irr \bfx \,  \zeta \hs d\bfx,  \;\;\;\;\;\;\;\;   
	I (\zeta)  = \irr |\bfx |^2 \zeta \hs d \bfx.
\end{equation}
The flow in the plane induced by a compactly supported vorticity $\zeta$ has a velocity field 
that scales with $\Gamma |\bfx|^{-1}$ as $|\bfx| \rightarrow \infty$.  While its energy, linear momentum
and angular momentum are therefore divergent, the invariants 
$H(\zeta), B(\zeta), I(\zeta)$ represent the finite parts of those quantities  \cite{lamb}.   
Throughout we shall refer to $H$ and $I$ simply as the energy and angular impulse.       

A point vortex dynamics occurs when the vorticity distribution is sharply concentrated at $N$ points:
\begin{equation}   \label{point-vorticity} 
	\zeta^N(\bfx, t) = \sum_{i=1}^N \gamma_i \delta (  \bfx - \bfx_i(t) ) 
\end{equation}
where $\gamma_i$ is the vortex strength (or circulation)  at the point $\bfx_i$, 
the location of the $i^{th}$ vortex;  $\delta(\bfx)$ is the unit delta function on $\mathbb R^2$.       
The Euler equations then collapse to advection of the point vortices by the induced flow, whose streamfunction is
\begin{equation}   \label{point-streamfn} 
	\psi^N(\bfx, t) = -\frac{1}{2 \pi} \sum_{i=1}^N \gamma_i \log |\bfx - \bfx_i(t) |   \, .
\end{equation}
This classical reduction of the continuum equations to a finite-dimensional
dynamical system for the point vortex locations ignores the self-induced velocity field of each 
vortex  \cite{newton}.   While intuitively evident, this desingularization can be justified analytically 
in the point vortex limit of the continuum dynamics under some restrictions \cite{marchioro-pulvirenti}.       
 The point vortex dynamics may be expressed as a canonical Hamiltonian system  in the variables, 
$q_i = \sqrt{\gamma_i} x_i, \; p_i = \sqrt{\gamma_i} y_i$, after rewriting the desingularized energy,  
\begin{equation} \label{pvham}
	H^N (\bfx_1, \dots, \bfx_N)= - \frac{1}{4 \pi}  \sum_{ i , j = 1, i \neq j}^N  \gamma_i \gamma_j \log |\bfx_i - \bfx_j|   \, , 
\end{equation}
as a function $H^N(q_1, p_1, \ldots , q_n, p_n)$ \cite{marchioro-pulvirenti, newton}.      

Our interest centers exclusively on systems of identical vortices.     
Accordingly, we set $\gamma_i=1/N$, thereby normalizing the total circulation to unity.     
To give the governing equations a Hamiltonian formulation with respect to the spatial coordinates 
$\bfx_i = (x_i, y_i)$ themselves, we employ a rescaled Poisson bracket,  namely,   
\begin{equation}  \label{poisson-bracket}  
\{ F, G \} \, =  \, N \sum_{i=1}^N [ \, F , G \, ]_{(x_i, y_i)}   \, ; 
\end{equation}     
$F$ and $G$ denote generic functions on the phase space $\mathbb R^{2N}$ of the system,
and the planar bracket acts on their $\bfx_i$ variables separately.    
The point vortex dynamics is then equivalent to the family of identities: 
\[
\frac{d F}{dt} \, =\,  \{ F, H^N \}    
    \;\;\;\;   \text{   for all (smooth) real-valued functions } \; F = F(\bfx_1, \ldots \bfx_n) \, .   
\]  
Besides $H^N$ and $\Gamma^N=1$, this Hamiltonian dynamics conserves the center of vorticity and
the angular impulse, respectively,
\[
B^N = \frac{1}{N} \sum_{i=1}^N \bfx_i \, = \, 0 \, , \;\;\;\;\;\;    I^N = \frac{1}{N} \sum_{i=1}^N | \bfx_i |^2   \, ,  
\]
after centering the vortex system at the origin.    

Statistical equilibrium theory invokes the ergodic hypothesis and considers invariant probability distributions 
on the phase space  constrained by the known invariants \cite{balian,tuckerman}.
   In particular, the canonical Gibbs density, 
\begin{equation}    \label{gibbs}
\rho^N (\bfx_1, \ldots , \bfx_N) \, = \, \exp \left(  - N \beta H^N - N \alpha I^N -  \phi(\beta,\alpha,N)   \right)  \, ,   
\end{equation}  
is parameterized by constants $\beta$ and $\alpha$ conjugate to the ensemble mean values of 
energy and  angular impulse,  respectively,    
$\la H^N \ra = E$ and $ \la I^N \ra = L^2 $;  $\phi(\beta,\alpha,N)$ normalizes the probability density $\rho^N$.    
Here and throughout our subsequent discussion, the angle brackets $\la \cdot \ra$ denote average, or expectation,
with respect to a specified density; in this case $\rho^N$.
The invariant  $B^N$ is ignorable, as the centering constraint $\la B^N \ra =0$ is automatically satisfied.    

In (\ref{gibbs})  the canonical parameters  are scaled by $N$  in anticipation of the
appropriate continuum limit, in which $E$ and $L^2$ are fixed and finite as $N \rightarrow \infty$.
In that scaled limit  the theory of large deviations characterizes the 
behavior of the empirical measure $\zeta^N$ in (\ref{point-vorticity}), that is,  
the spatial density of point vortices \cite{eyink-spohn}.   Specifically,  $\zeta^N$  
tends in the weak sense of measures on $\mathbb R^2$ to the maximizer $\zeta$ of the entropic functional 
\[
S_{\beta,\alpha} [\zeta] \, = \,  -  \irr \zeta(\bfx) \log \zeta(\bfx) d \bfx  \, - \, \beta H(\zeta) - \alpha I(\zeta)  \;\;\;\;\;   
\]
over probability densities $\zeta(\bf x)$ on $\mathbb R^2$.   The same large deviation principle quantifies
the likelihood of fluctuations away from the most probable density $\zeta( \bf x) $ for large $N$: the
likelihood of another state $\tilde{\zeta}$ is exponentially small in $N$ with a rate functional equal to  
 $S_{\beta,\alpha}[\zeta] - S_{\beta,\alpha}[\tilde{\zeta}] \;  $.  The relevant
 large deviation theory is explained in  \cite{ellis-leshouches} and its application to point vortex ensembles 
 is summarized in \cite{turk-leshouches}.        

The optimality conditions for the equilibrium state $\zeta $  imply that its streamfunction satisfies 
the mean-field equation
\begin{equation}   \label{meanfield}          
 - \Delta \psi = e^{ -\beta \psi - \alpha | \bfx |^2 - \mu}    \, ; 
\end{equation}  
$\mu$ being the multiplier for the normalization, $\int \zeta \, d \bfx =1$.    Thus the most probable state 
in the continuum limit with finite energy and angular impulse is a special steady solution 
of the Euler equations, having an exponential dependence of vorticity on streamfunction.   That special dependence
in the macrostate is a direct consequence of the modeling assumption that the microstate consists of many
identical point vortices \cite{marchioro-pulvirenti, newton}.

It is often advantageous to associate these equilibrium flows with the energy and angular impulse values that 
they attain, and to parametrize branches of solutions to the mean-field equation (\ref{meanfield})  by 
$E$ and $L^2$ rather than $\beta$ and $\alpha$.   
Of course, this change of perspective is realized by replacing  the canonical ensemble by the
microcanonical ensemble.      
Microcanonical equilibrium states are constrained maximizers  of entropy:      
\begin{eqnarray} \label{maxent}
	\mbox{ maximize  } \;  -  \irr \zeta(\bfx) \log \zeta (\bfx) d \bfx \;\;\;\;\;\;\;\;\; \mbox{ subject to }   \\
	\frac{1}{2} \irr \psi(\bfx) \zeta(\bfx) d \bfx = E, \;\;\;\;   \irr |\bfx|^2 \zeta(\bfx) d \bfx = L^2,   \;\;\;\;
	\irr \zeta(\bfx) d\bfx = 1  \, ;  \nonumber      
\end{eqnarray}  
$\beta, \alpha, \mu$ are then the  Lagrange multipliers associated with the equality constraints,
and the maximizer satisfies the mean-field equation \eqref{meanfield}.    
A large deviation analysis similar to that for canonical equilibria is applicable to microcanonical equilibria 
\cite{turk-leshouches}.       
An iterative algorithm exists to solve the constrained optimization problem (\ref{maxent})
and thus to compute branches of equilibrium states  \cite{turk-whitaker}.   Using this algorithm it is possible
to establish that  the maximum entropy is a concave function of the constraint values $E,L^2$, and hence
that the canonical and microcanonical formulations are equivalent in this particular problem;  for vortex systems
under other conditions the equivalence of ensembles may break down when $\beta <0$ \cite{kiessling-lebowitz,EHT}.

\begin{figure}[bt]    
	\centering
	\includegraphics[width=7cm]{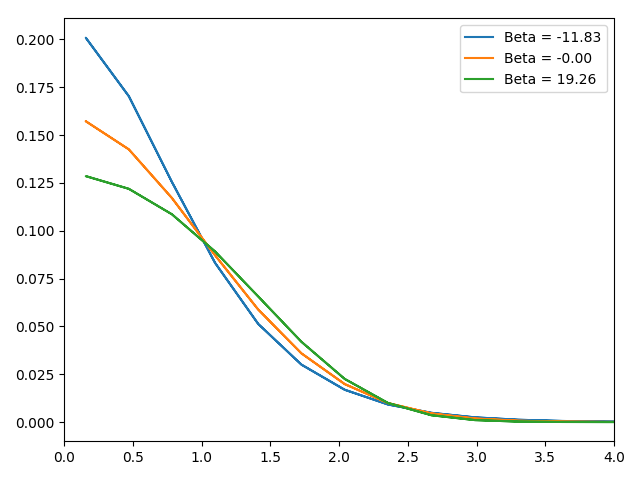}
	\caption{Radial profiles of vorticity for representative equilibrium states with 
	 $\beta<0, \; \beta=0, \; \beta>0$.  }
        \label{equil-densities}  
\end{figure}

These equilibrium states are axisymmetric, stable, macrovortices
with radially decreasing vorticity.    For $\beta = 0$ they are Gaussian densities,
while for $\beta > 0$ they are less concentrated, and for $\beta <0$ they are more concentrated
at the origin than a Gaussian.    Figure \ref{equil-densities} displays three representative radial profiles
of the vorticity $\zeta(r)$.    

\section{Optimal Closure Method}

When we endeavor to construct a nonequilibrium theory of macroscopic states, and closed evolution equations
for those states, we necessarily include observables in the macroscopic description  that are not exact
invariants.    Accordingly, each choice of macroscopic observables leads to a corresponding nonequilibrium reduced
model.   This arbitrariness can be circumvented only when the system under study has a
strong separation of time scales, so that its slow variables can form the macroscopic description, and its
fast variables can be relegated to a statistical treatment.   Even in this case a unique reduction is only achieved
in the limit as the separation of time scales becomes infinite.    The current paper therefore adopts a different perspective.
Our approach to model reduction accepts that some choice of resolved, macroscopic observables must be made,
and that this choice is not unique.   Given a chosen set of observables ---  presumably a reasonable choice from
the standpoint of a desired reduced model --- the mathematical problem then becomes how to devise a 
suitable closure in terms of those observables.   

  We approach this problem using a combination of statistical
modeling and information theory.    Namely, we construct a parametric family of ``trial" probability densities on the phase
space of the microscopic dynamics (assumed to be deterministic and Hamiltonian),  having
parameters that are in a one-to-one correspondence with the mean values of the chosen observables.    Within 
the reduced statistical model defined by that parametric family we attempt to follow the propagation of probability 
induced by the underlying microscopic dynamics.    To do so we quantify the deviation between the exactly
propagated density and any feasible evolution of the model density,  using a metric derived from the relative entropy, 
or Kullback-Leibler divergence, between the exact and model densities.     Dynamical closure within the reduced model is then 
determined by minimizing this metric over feasible paths in the parameter space of the model.
In this way, we obtain a statistically consistent closure, in which macrostates evolve so as to 
minimize the rate of information loss due to model reduction.   

This optimal closure method of model reduction was introduced by one of the authors in \cite{turk-JSP}.
In that paper a weighted metric is considered, the weights being adjustable parameters in the closure.
Subsequent studies have shown, however, that these adjustable weights are unnecessary, and that an intrinsic
optimal closure is achieved by setting all weights equal to unity \cite{kleeman,TCT,TT2}.   

In the remainder of this section we summarize the optimal closure method in the general context of reducing 
a Hamiltonian dynamics onto a finite set of linearly independent observables, $A_1, \ldots, A_m$.   Let $z$ 
denote a generic point in the phase space $R^{2N}$;  for instance, in the point vortex system, $z = (\bfx_1, \ldots , \bfx_N)$.     
For ``trial" probability densities on the phase space we take quasi-canonical densities built from the chosen
observables;  namely,
\begin{equation}   \label{trial-density}
 \rhotilde(z; \xi) \, = \, \exp \left( \, \sum_{k=1}^m \xi_k A_k(z) - \beta H(z) - \phi(\xi, \beta )  \,    \right)  \, , 
\end{equation}
where $\xi = (\xi_1, \ldots, \xi_m)$ is the parameter vector conjugate to the resolved vector $A=(A_1, \ldots , A_m)$;
$\phi(\xi, \beta )$ normalizes the probability.    The probability densities (\ref{trial-density}) constitute an 
exponential family with natural parameters $\xi \in \mathbb R^m$ and $-\beta \in \mathbb R$ 
\cite{casella-berger,wasserman}.   We are interested
in modeling the relaxation of macrostates toward statistical equilibrium, and so we slave $\beta$ to $\xi$ by
imposing the conservation of energy, namely, that $\la H \ra = E$, where $E$ is the equilibrium energy.   Here the expectation is taken with respect to the trial density $\rhotilde$.    We defer until the next section any specification of the vector  observable
$A$, except to say that $A(z)$ represents some coarse-graining of the microstate $z$.    

The exact propagation of probability on the phase space  $R^{2N}$ is expressible formally as 
\[
\rho(t) = e^{- (t-t_0) L } \rho(t_0) ,   \;\;\;\;\;\;\;\; \mbox{ where } \;\;   L = \{ \cdot , H \}   \, .  
\]
Indeed, this statement is equivalent to the Liouville equation, 
\[
\d_t \rho \, + \, \{ \rho , H \} \, = \, 0 \, ,   
\]
which is the foundation of statistical mechanics \cite{balescu,tuckerman,zwanzig}.    
Our closure procedure is based on submitting trial densities $\rhotilde(\cdot; \xi(t))$ to the Liouville equation and 
minimizing the mean-square of the resulting residual over parameter paths $\xi(t)$.    
Equivalently, our closure principle can be viewed in information-theoretic terms by
analyzing the Kullback-Leibler divergence between an exactly propagated density and a feasible model density 
\cite{kullback,cover-thomas}.    
 Namely, let $\rhotilde(t) = \rho(t)$ at time $t$, and for a short interval of time $\Delta t$ consider the  
incremental relative entropy,  
\begin{eqnarray}    \label{KL}   
D_{KL} ( \, e^{- \Delta t L} \rho(t) \, \| \, \rhotilde(t + \Delta t) \, )  &=&   \nonumber 
   \int_{R^{2N}} e^{- \Delta t L} \rho(t) \, \log \frac{ e^{- \Delta t L} \rho(t) } {  \rhotilde(t + \Delta t) } dz \\ 
   &=&  \frac{ (\Delta t)^2}{2}   \left\la \left[  \, (  \d_t + L ) \log \rhotilde(\cdot ; \xi(t) )   \right]^2 \right\ra  \; +  \; 
                    o ( \, ( \Delta t)^2 \, ) \, ,  \nonumber 
\end{eqnarray}  
expanding in a Taylor series in $\Delta t$;  
the expectation in the second equality is taken with respect to $\rhotilde(\cdot ; \xi(t))$.    
The leading order term in
this expansion of the relative entropy is an intrinsic metric on the discrepancy
between the exact and model densities.  In view of this calculation, we define
the  Liouville residual to be 
\begin{equation}  \label{L-residual}   
R \, = \, R (\xi, \dot{\xi} )  = \, (  \d_t + L ) \log \rhotilde(\cdot ; \xi(t) )   \, , 
\end{equation} 
on any feasible path $\xi(t)$, and we quantify the local lack-of-fit of the reduced model by 
\begin{equation}  \label{lof-lagrangian} 
\L (\xi, \dot{\xi} ) = \frac{1}{2} \la R^2 \ra \,  .     
\end{equation} 
Our optimal closure minimizes the time integral of  $\L (\xi, \dot{\xi} )$ over feasible parameter paths.    
These extremal paths realize the slowest rate of information loss, or
entropy production,  within the reduced model, and thus represent the best fit of the trial  
densities to the Liouville equation.        

The optimality conditions for the closure are conveniently derived by applying
Hamilton-Jacobi theory \cite{gelfand-fomin,evans,liberzon}.
That is, we view $\L$ as a Lagrangian function and introduce its value function (or action integral), 
\begin{equation}   \label{value-fn}
v(\bar{\xi}, \bar{t} ) \, = \, \min_{\xi(t), \xi(0) = \bar{\xi}} \int_0^{\bar{t} } \L (\xi, \dot{\xi} ) \, dt \, .
\end{equation}    
To the lack-of-fit Lagrangian we associate a lack-of-fit Hamiltonian 
(not to be confused with that for the microscopic dynamics)
and conjugate variables, namely, 
 \begin{equation}   \label{lof-hamiltonian}
\H(\xi, \pi ) = \max_{\dot{\xi}} \; \sum_{k=1}^m \pi_k \dot{\xi}_k \, - \, \L (\xi, \dot{\xi} ) \, , \;\;\;\;\;\;
   \;\;\;\; \;\;   \pi_k = \frac{\d \L}{\d \dot{\xi}_k }    \, .
\end{equation}   
The value function (\ref{value-fn}) satisfies the Hamilton-Jacobi equation
\begin{equation}   \label{HJ-eqn}
\frac{\d v}{\d t} \, + \, \H (\xi , - \frac{\d v}{\d \xi} ) \, = \, 0 \, ,  \;\;\;\;\;\; \mbox{ with }  \;\;  v(\xi,0) = 0 \, .  
\end{equation}   
Along extremals, $\,   \pi_k =  - \d v /  \d\xi_k  \, $, and hence we obtain the 
reduced equations
\begin{equation}  \label{closed-reduced-eqn} 
\frac{\d \L}{\d \dot{\xi}_k } ( \xi , \dot{\xi} )  =  - \frac{\d v}{\d \xi_k}(\xi)    \;\;\;\;\;\;\;\; ( \, k = 1, \ldots m \, ) \, ,  
 \end{equation}   
 a closed system of $m$ ordinary differential equations of first order in $\xi(t) \in \mathbb R^m$.   
Our optimal closure approximates the relaxation of the system from an initial statistical state $\rhotilde(\cdot,; \xi_0)$
to statistical equilibrium (for which $\xi =0$) by the path of trial densities  $\rhotilde(\cdot; \xi(t))$ 
determined by the solution $\xi(t)$ of (\ref{closed-reduced-eqn}) satisfying the initial condition $\xi(0) = \xi_0$.  
 
 The Liouville residual for a trial density (\ref{trial-density}), in which $\beta(\xi)$ is 
 determined by the mean energy constraint $\la H \ra = E$, is given by   
 \[
 R \, = \, \sum_{k=1}^m \dot{\xi}_k  \, Q_H ( \, A_i - \la A_k \ra \, ) \, + \, \xi_k \{ A_k , H \} \, , 
 \] 
 where $Q_H$ denotes the complementary orthogonal projection onto the energy shell, that is,
 \[
 Q_H F = F - P_H F \, , \;\;\;\;\;\; \mbox{ with } \;\; P_H F = \frac{ \la F ( H-E) \ra }{ \la (H-E)^2 \ra } (H-E)  \, .   
 \]
 From this formula it is evident that $\la R \ra =0, \;\; \la R (H -E) \ra = 0$, and hence that the dynamical
 diagnostic $R$ is analogous to the score variable for a parametric statistical model \cite{wasserman}.  
 In turn,  its mean-square $\la R^2 \ra$, which defines the lack-of-fit (\ref{lof-lagrangian}), is a dynamical 
 analogue to the Fisher information \cite{wasserman}.    
 The meaning of the conjugate variables $\pi_k$ is revealed by the calculation, 
  \[
 \pi_k = \frac{\d \L}{\d \dot{\xi}_k } = \la Q_H ( A_k - \la A_k \ra ) \, R \ra = \la A_k R \ra \, ,   
 \]
 which uses the fact that $\L (\xi , \dot{\xi})$ depends quadratically on $\dot{\xi}$.   
 Taking the moment with respect to $A_k$ of the Liouville equation yields 
 \begin{equation}   \label{moment-eqn}  
 \frac{ d}{dt} \la A_k \ra \, = \, \la \{ A_k , H \} \ra \, + \la A_k  R \ra \, .    
 \end{equation}  
 Thus, $\pi_k$ is precisely the residual in the moment equation for $A_k$, due to replacing the exact density
 by the trial density.  Accordingly, $\pi_k$ may be interpreted thermodynamically as the
 irreversible flux of the mean $\la A_k \ra$.    The Hamilton-Jacobi analysis reveals that the flux vector, $\pi$, 
 is the minus $\xi$-gradient of the value function $v(\xi, t)$.  Therefore $v$ has the thermodynamic 
 interpretation of a dissipation potential in the reduced equations \cite{degroot-mazur}.       
 
 A fuller explanation of this optimal closure is
 given in \cite{turk-JSP}, where it is termed the ``nonstationary closure," since the value function is
 time-dependent.    In this closure the dissipation vanishes at $t=0$ when the initial nonequilibrium
 statistical state coincides with a trial density;  for large $t$, it approaches the associated ``stationary closure," 
 for which the value function is time-independent.        
 
 Practical implementation of this model reduction procedure is inhibited by the difficulty inherent in
 solving the Hamilton-Jacobi equation (\ref{HJ-eqn}), which is not analytically tractable apart from very
 special (completely separable) cases.   
 This impediment may be overcome by resorting to numerical optimization methods, such as
 are available in the optimal control literature; for instance, this approach is implemented in \cite{TT2}.    
 In the current paper, however, we instead restrict
 our attention to near-equilibrium relaxations, which allows us to linearize the optimization
 theory around equilibrium and thereby obtain a linear irreversible closure.      
 Under that approximation the Hamilton-Jacobi equation becomes a matrix Riccati equation,
 whose solution may be identified with the matrix of transport coefficients that govern the dissipative reduced
 dynamics.     Rather than develop this approximation in the abstract, we return in the next section to point 
 vortex systems and elaborate the optimal closure in that setting.

\section{Nonequilibrium Mean-Field Theory of Vortex Systems}

We coarse-grain the $N$-vortex dynamics, for large $N$, in terms of finitely-many spatial moments of the
vorticity defined by the functions $A_k(\bfx), \; k =1\, \ldots, m$.  
The macrostate is then the ensemble mean of the vector observable 
$A^N  \in \mathbb R^m$ having components 
\begin{equation}  
A_k^N(\bfx_1, \ldots , \bfx_N) = \frac{1}{N} \sum_{i=1}^N A_k (\bfx_i)     \, . 
\end{equation}  
(Abusing the notation of the preceding section, henceforth  $A_k^N$ denotes the observable
on phase space, while $A_k$ denotes the spatial moment on physical space.)  
For instance, when we model the axisymmetrization of a distorted vortex ensemble centered at the
origin (so that the first-order moments vanish and are ignorable), we take 
 $A_1 = x^2-y^2, A_2 = 2xy$,  along with $A_3= x^2+y^2$, which determines the angular impulse invariant. 
Heuristically, the macrostate, $a = \la A \ra \in \mathbb R^3$,  describes the effective ``size", ``eccentricity" and ``inclination" 
of the evolving vortex ensemble conceived as an elliptical structure.    

In this section we include the angular impulse in the resolved vector, for the sake of clarity while elaborating 
the mean-field theory.  In the later sections, we treat the angular impulse as an invariant, like the
energy, and include only non-conserved resolved variables in $A$.

\subsection{Mean-field Ansatz} 

Motivated by the established fact that the equilibrium theory of vortex systems yields an asymptotically
exact mean-field theory in the appropriate continuum limit, we base our nonequilibrium theory on 
the following trial densities on $\mathbb R^{2N}$:    
\begin{equation} \label{trial-densities-MF}
  \rhotilde_N (\bfx_1, \ldots, \bfx_N ; \beta, \xi) =   \prod_{i=1}^N \zetatilde (\bfx_i; \beta, \xi)  
\end{equation}
where $\zetatilde(\bfx; \beta, \xi)$ is the probability density on $\mathbb R^2$ defined by the mean-field equation
\begin{equation}    \label{noneq-MF} 
    \zetatilde = \, -  \Delta \psitilde \, = \,  e^ {  \sum_{k=1}^m \xi_k A_k   - \beta \psitilde   - \mu }  \, .  
 \end{equation}   
That is, our statistical model of the $N$-vortex system assigns independent random locations, $\bfx_i$, to
each of the point vortices, distributed according to $\zetatilde(\bfx)$.   This spatial density is parameterized by 
$\beta$, conjugate to the energy invariant, $E$,  and $\xi_1, \ldots , \xi_m$ conjugate to the moments, 
$A_1(\bfx), \ldots , A_m(\bfx)$ that define the coarse-graining; $\mu = \mu(\beta,\xi)$ normalizes $\zetatilde$.
   In the limit as $N \rightarrow \infty$, the 
law of large numbers implies that the empirical density $\zeta^N(\bfx)$, defined in (\ref{point-vorticity}), tends
to $\zetatilde(\bfx)$ in the weak sense of measures;  in turn, the smoothing properties of the Poisson equation \cite{evans}
guarantee that the corresponding streamfunction $\psi^N(\bfx)$, defined in (\ref{point-streamfn}), tends to $\psitilde(\bfx)$, 
as do its first derivatives and hence the corresponding velocity field.    

The trial densities (\ref{trial-densities-MF}) are constructed from solutions $\psitilde(\bfx)$ of a nonlinear 
elliptic partial differential equation (\ref{noneq-MF}),  which depends parametrically on $\beta$ and $\xi$.   
Alternatively, the marginal density, $\zetatilde(\bfx)$, is  the solution of the maximum entropy problem
(\ref{maxent}) in which the single constraint on angular impulse is replaced by the family of constraints:
\[
\int_{\mathbb R^2} A_k (\bfx) \zeta (\bfx) \, d \bfx \, = \, a_k    \;\;\;\;\;\;\;\; \mbox { for } \;\; k = 1, \ldots, m .  
\]
The constraint values, $a_k$ and $E$, then parameterize the maximizers $\zetatilde(\bfx)$ rather
than the associated multipliers $-\xi_k$ and $\beta$.  Holding the energy constant, $\beta$ is slaved
to $\xi$, and hence $\xi $ is determined by the macrostate vector $a$.   The correspondence between $\xi$
and $a$ is one-to-one in a neighborhood of equilibrium, as our subsequent analysis will show.  

We submit this mean-field trial density to the Liouville equation governed by 
the Hamiltonian  $H^N$ displayed in \eqref{pvham} (with $\gamma_i = 1/N$), 
and the Poisson bracket (\ref{poisson-bracket}).  The  scaled residual is 
\begin{eqnarray}   \label{residual-full-MF} 
     R &=& \frac{1}{N} (\, \partial_t + \pbpv{\cdot}\, ) \log \rhotilde_N (\bfx_1, \ldots, \bfx_N; \beta,\xi)  \\
	& =& \frac{1}{N} \sum_{i=1}^N ( \, \partial_t + [ \, \cdot , \psi^N_i(\bfx_i) \,  ] \, ) 
	                      \left(  \sum_{k=1}^m \xi_k A_k(\bfx_i) \, - \, \beta \psitilde(\bfx_i) \, - \mu(\beta,\xi) \,  \right)   \nonumber \\
	&=& \frac{1}{N} \sum_{i=1}^N \;\; \sum_{k=1}^m \dot \xi_k ( A_k(\bfx_i) - \la A_k \ra) +  \xi_k  [ A_k , \psi^N_i ] (\bfx_i) \nonumber \\  
	&  &   \;\;\;\;\;\;\;\; \;\;\;\;       -   \dot \beta (\psitilde(\bfx_i) - \la \psitilde \ra) 
	                                          -   \beta  \partial_t (\psitilde(\bfx_i) - \la \psitilde \ra )   
	                                          -   \beta [ \psi^N_i , \psitilde ] (\bfx_i)     \, .   \nonumber      	  
\end{eqnarray}
This calculation makes use of the desingularized streamfunction,    
\[
\psi^N_i (\bfx) =  -  \frac{1}{2\pi N} \sum_{j \neq i }  \log | \bfx - \bfx_j | \, ,   
\]
and the identity, $ \{ F(\bfx_i) , H^N (\bfx_1, \ldots , \bfx_n) \} = [ F , \psi^N_i ] (\bfx_i)$.   
In (\ref{residual-full-MF}), and in the analysis to follow, 
$\la \cdot \ra$ denotes expectation with respect to the density $\zetatilde$.    
We notice that the mean-field ansatz produces  two terms that do not appear in the residual of
quasi-canonical trial densities outlined in the preceding section.    The  term, $ \beta  \partial_t (\psitilde - \la \psitilde \ra ) $, 
accounts for the implicit dependence of the streamfunction $\psitilde$ on the parameters $\xi(t), \beta(t)$.   
The term, $  \beta [ \psi^N_i , \psitilde ]  $, vanishes in the continuum limit.     
Both these terms reflect the relative simplicity of the mean-field ansatz, which for finite $N$ imposes 
the product form (\ref{trial-densities-MF}) that the quasi-canonical density achieves only in the limit as $N \rightarrow \infty$.

It is evident from (\ref{residual-full-MF}) that, in the continuum limit as $N \rightarrow \infty$, the appropriate
continuum residual on which to base a mean-field optimal
closure is  
\begin{equation}   \label{residual-MF}   
	\tilde{R}(\bfx; \xi, \dot \xi, \beta, \dot \beta)  = 
	  \sum_{k=1}^m \dot \xi_k ( A_k - \la A _k\ra) + \xi_k  [ A_k, \psitilde]  - \dot \beta (\psitilde - 2E)  - \beta \partial_t \psitilde,
\end{equation}
noticing that $\la \psitilde \ra = 2E$, which is constant in time.    
This mean-field version of the Liouville residual is derived entirely from $\zetatilde$.  In fact, it
 may be viewed as simply the residual in the Euler equations  of the ansatz (\ref{noneq-MF}), that is,   
\begin{equation}
	\tilde{R} =  \left(\partial_t + \left[ \, \cdot, \tilde \psi \,  \right] \right) \, \log \tilde \zeta.
\end{equation}
In this light, the derivation of $\tilde{R}$ from the mean-field ansatz may be viewed 
as a justification on information-theoretic grounds for using $\L = \la \tilde{R}^2 \ra /2 $ as the 
cost function for an optimal closure formulated in physical space in terms of
quasi-equilibrium vorticity fields (\ref{noneq-MF}).     The reduced model is therefore an optimal closure with respect to
the Euler equations themselves, in which  the underlying point vortex dynamics and its associated entropy functional 
are used to deduce the trial densities and the lack-of-fit metric appropriate to the continuum limit.

\subsection{Near-Equilibrium Linearization}

Next we linearize the optimal closure theory formulated in the preceding subsection around a given equilibrium
state, $\zeta_{eq}$, with corresponding streamfunction, $\psi_{eq}$.     Relaxation of an initial disturbed macrostate
toward this equilibrium macrostate conserves the energy, $E$, and angular impulse, $L^2$, and thus the  
near-equilibrium linearization of the optimal closure is required to
respect those invariants up to linear order in the perturbation.    The mean-field macrostate is 
\begin{equation}    \label{noneq-MF-again} 
    \zetatilde = \, -  \Delta \psitilde \, = \,  e^ {  \sum_{k=1}^m \xi_k A_k  -\alpha |\bfx|^2   - \beta \psitilde   - \mu }  \, ,   
 \end{equation} 
and by assumption the parameter vector, $\xi \in \mathbb R^m$, is near the origin.   
(The angular impulse term is no longer included in the resolved vector, since it is an invariant.)
Denoting perturbations from
equilibrium by primes, so that $\alpha = \alpha_{eq} - \alpha'$, $\beta = \beta_{eq} - \beta'$ and 
$\zetatilde = \zeta_{eq} + \zeta'$, $\psitilde = \psi_{eq} + \psi'$, we have 
\[
\zeta' \, = \, \left[ \, \sum_{k=1}^m \xi_k A_k + \alpha' ( |\bfx|^2 - L^2) + \beta' (\psi_{eq} -2E) - \beta_{eq} \psi'  \, \right] \, \zeta_{eq} 
   \; + \; O( |\xi|^2).
\]
Consequently the perturbation streamfunction satisfies 
\begin{equation} \label{pde-psiprime}   
   ( \, - \Delta + \beta_{eq} \zeta_{eq} \, ) \psi'   \,   = \, 
            \left[ \, \sum_{k=1}^m \xi_k A_k + \alpha' ( |\bfx|^2 - L^2) + \beta' (\psi_{eq} -2E)   \, \right] \, \zeta_{eq}   \, . 
\end{equation}    
This elliptic equation along with the boundary condition,   $\lim_{  | \bfx | \rightarrow \infty}      \psi' ( \bfx )  =  0  \,$, 
determines the solution $\psi'$, which depends linearly on $\xi, \alpha', \beta'$.   In turn, $\alpha', \beta'$ are determined
as linear functions of the vector $\xi$ by solving the linearized energy and angular impulse constraints,  
\[
\irr \psi_{eq} \, \zeta' \, d \bfx =0 ,   \;\;\;\;\;\;    \irr |\bfx|^2 \zeta' d \bfx = 0.
\]
In summary, the linearized mean-field ansatz together with the linearized invariants yields the linear expressions 
\begin{equation}   \label{sensitivities}   
	\alpha' = \sum_{k=1}^m \alpha_k  \xi_k \, , \;\;\;\; 
	\beta' = \sum_{k=1}^m  \beta_k  \xi_k  \, , \;\;\;\;
	\psi' = \sum_{k=1}^m  \psi_k \xi_k  \, , 
\end{equation}
in which the sensitivities, denoted by  $\alpha_k$, $\beta_k$, $\psi_k$,  to the perturbation $\xi_k$ are calculable 
from the given equilibrium state $\zeta_{eq}$.   
We refrain from displaying explicitly the coefficient matrices involved; full details are given in \cite{maack-thesis}.

The Liouville residual is, to leading order,
\begin{equation} \label{nelr}
	R = \dot{\xi}^T \mathbf U + \xi^T \mathbf V.
\end{equation}
where we define the vectors $\mathbf U$ and $\mathbf V$ by
\begin{align}  \label{UV} 
	U_k &= A_k + \alpha_k(|\bfx|^2 - L^2) + \beta_k(\psi_{eq} - 2E) - \beta_{eq}\psi_k,  \\
	V_k &= [ \, A_k, \psi_{eq} \, ] - \alpha_{eq}[ \, |\bfx|^2, \psi_k \, ]. \nonumber   
\end{align}
Squaring and taking the expectation we obtain the quadratic lack-of-fit Lagrangian, 
\begin{equation}  \label{lagrangian-MF} 
    2 \L(\xi, \dot{\xi} ) \, = \,  \lex R^2 \rex_{eq} = 
        \dot{\xi} ^T \lex \mathbf U \mathbf U^T \rex_{eq} \dot \xi + 2\xi^T \lex \mathbf V \mathbf U^T \rex_{eq} \dot{\xi} 
                                                + \xi^T \lex \mathbf V \mathbf V^T \rex_{eq} \xi   \,  ,
\end{equation}    
whose matrices 
\begin{equation} \label{matrices}
	C = \lex \mathbf U \mathbf U^T \rex_{eq} \, ,  \;\;\;\; 
	J = \lex \mathbf V \mathbf U^T \rex_{eq}  \, ,  \;\;\;\;  
	K = \lex \mathbf V \mathbf V^T \rex_{eq}  \, , 
\end{equation}
are computed from the given equilibrium state $\zeta_{eq}$.   

The optimal closure thus becomes a quadratic programming problem and 
its Hamilton-Jacobi equation (\ref{HJ-eqn})  becomes a Riccati matrix equation for the $m \times m$ matrix, $M(t)$, 
that defines the quadratic value function, $v(\xi,t) = \xi^T M(t) \xi / 2 \, $.   
Specifically,  $M(t)$ is  the solution of the matrix initial value problem
\begin{equation}   \label{riccati}
	\dot M + M C^{-1} M + J C^{-1} M - M C^{-1} J = D, \;\;\;\; \mbox{ with } \;\;  M(0) = 0,
\end{equation}
where $D = K + J C^{-1} J$.    Known properties of Riccati equations ensure that 
$M(t)$ is symmetric semi-definite, 
because $C $ is symmetric positive definite and $D$ is symmetric semi-definite \cite{liberzon}.    
The optimal parameter path $\xi(t)$  satisfies the linear system
\begin{equation} \label{reduced-eqn-MF}
	C  \, \frac{d \xi}{dt} \,  =  \, (\, J - M(t) \, ) \, \xi   \, ,       
\end{equation}
which is the linearization of (\ref{closed-reduced-eqn}).   
Since the macrostate vector $a$ and the parameter vector $\xi$ are related by
\[
	a (t) =  \irr  A (\bfx) \, \zetatilde(\bfx, t) \, d \bfx  \,  = \, C \xi (t) + O(|\xi|^2) \, , 
\]
the reduced equation (\ref{reduced-eqn-MF}) is equivalently a relaxation equation for the macrostate. 
This completes the optimal closure.   We emphasize that this linearized, near-equilibrium relaxation 
equation is constructed entirely from the given equilibrium mean-field state, $\zeta_{eq}$, and is free
of any adjustable constants, such as relaxation rates.   Moreover, the reduced dynamics is memoryless
in the macrostate $a(t)$, although it is non-autonomous due to the time dependence of $M(t)$.    

The remainder of the paper applies this optimal closure to two simplified models arising in  geophysical
fluid dynamics.   These models are straightforward generalizations of the two-dimensional Euler equations,
and the analysis given above extends to them without any fundamental changes.    We therefore concentrate
on displaying and interpreting the predictions of the closure, and testing those predictions against benchmarks 
computed by direct numerical simulations of the many-vortex systems.

\section{Axisymmetrization in Single-Layer Dynamics}

The single-layer quasi-geostrophic dynamics is governed by the  transport equation for
the potential vorticity, denoted by $q(\bfx,t)$:    
\begin{subequations} \label{singlelayer}
\begin{align}
	\partial_t q + [q, \psi] &= 0, \label{slvort} \\
	-\Delta \psi +   R_d^{-2} \psi &= q. \label{slstream}
\end{align}
\end{subequations}
We refer the reader to the literature on geophysical fluid dynamics for the meaning of this system,
and its derivation as an asymptotic model for a rotating shallow fluid layer in the limit of small
Rossby number \cite{pedlosky,salmon, bouchet-venaille}.  
This model involves a length scale, $R_d$, the Rossby deformation radius,
which characterizes the scale at which kinetic and potential energies balance.       
An ensemble composed of $N$ concentrations of potential vorticity, namely,   
\[
q(\bfx,t) = \sum_{i=1}^N \frac{1}{N}  \delta(\bfx - \bfx_i) 
\]
interacts with a screened potential; specifically, the Green function for (\ref{slstream})     
is the Bessel function $(1/2\pi) K_0( r/R_d)$.  The Euler equation is recovered in the limit as $R_d \rightarrow \infty$.  

The presence of a length scale in the interaction potential for the vortex system allows us to investigate
how the relaxation of an ensemble of vortices  depends on the ratio between $R_d$ 
and the average ``radius" of the ensemble, which thanks to the angular impulse constraint is $L$ for
an ensemble centered at the origin.      
To do so we study the evolution of ensembles of vortices  that are
initially distorted away from a given axisymmetric equilibrium state.    We measure this distortion by 
the second moments of the vorticity field.    Besides simplicity, the  justification for this choice is that
a centered Gaussian distribution is completely characterized by it second moments; and equilibrium macrostates
for $\beta =0$ are Gaussians.   
The moment of $|\bfx|^2 = x^2 + y^2$ being a conserved quantity, we declare the observables
\begin{equation} \label{eeslobs}
	A_1(\mathbf x) = x^2 - y^2, \hspace{5mm} A_2 (\mathbf x) = 2xy \, , 
\end{equation}
to be the resolved variables, noticing that $A_1(\bfx) , A_2(\bfx) $ and $|\bfx |^2$
form an orthogonal basis for the (centered) quadratics in $\bfx$ 
with respect to any axisymmetric density, and hence any equilibrium macrostate.       
The ensemble mean values, $a_1=\la A_1 \ra$, $a_2=\la A_2 \ra$, relax to zero as the
ensemble equilibrates.  We normalize the spatial scale by setting $L^2=2$, so that a 
symmetric Gaussian density has unit variance in both $x$ and $y$.    The ensemble evolution is initialized
by a centered Gaussian density with given $a_1, a_2$.      

In qualitative terms,  an elliptically distorted initial state develops spiral arms, which are entrained and sheared by
the large-scale rotation and eventually relax toward the axisymmetric equilibrium state.   Our severe coarse-graining
does not resolve the spiral arm structure, but instead fits trial densities to it 
that track its ``shape,"  quantified by the  mean observables, $\la A_1 \ra, \la A_2 \ra$ along with the
angular impulse  $\la | \bfx |^2 \ra$, and its ``concentration,"  controlled by the energy $\la \psi \ra /2$.

\begin{figure}[bt]
\begin{tabular}{cc}
	\includegraphics[width=7.2cm]{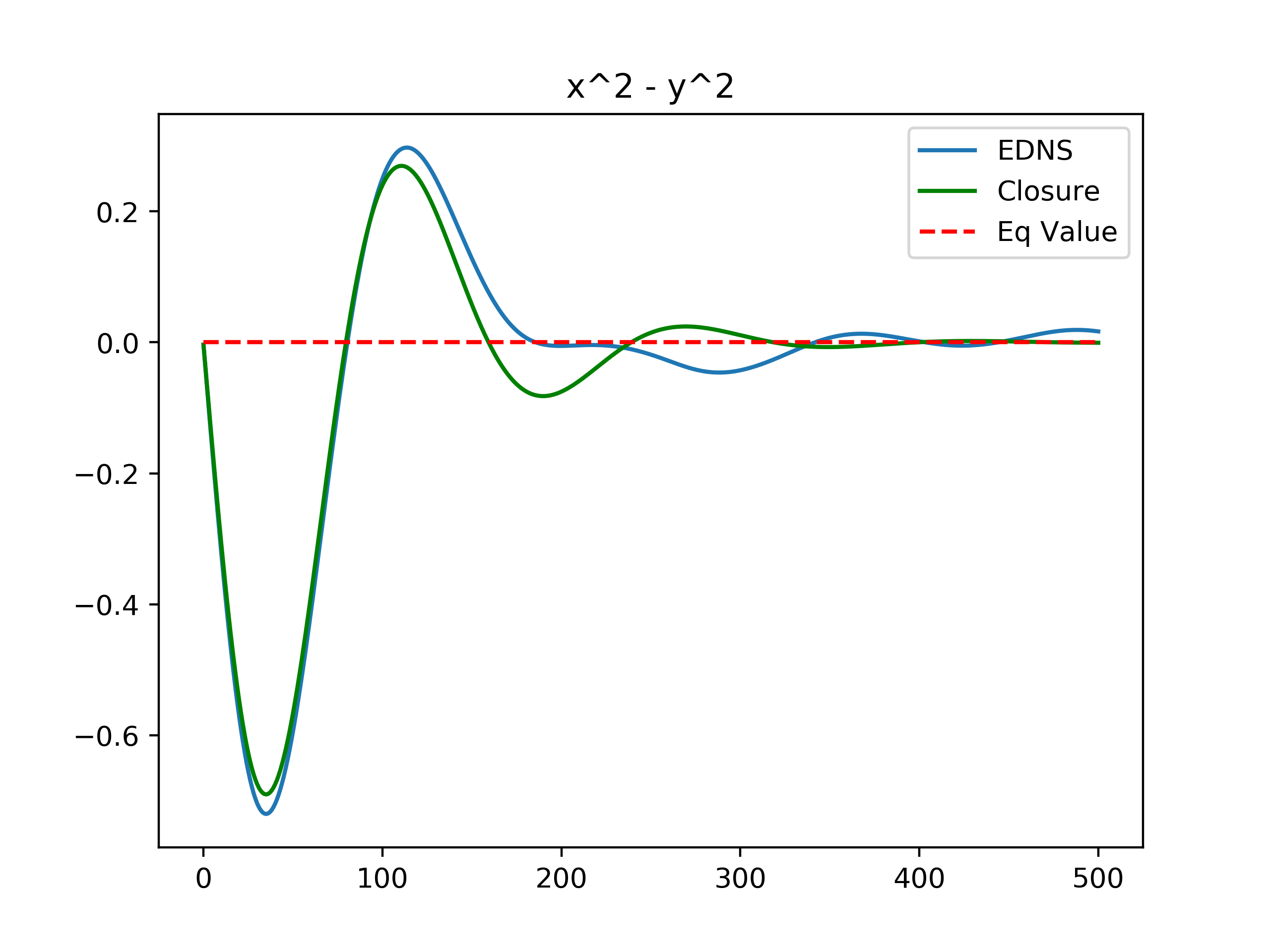} &
	\includegraphics[width=7.2cm]{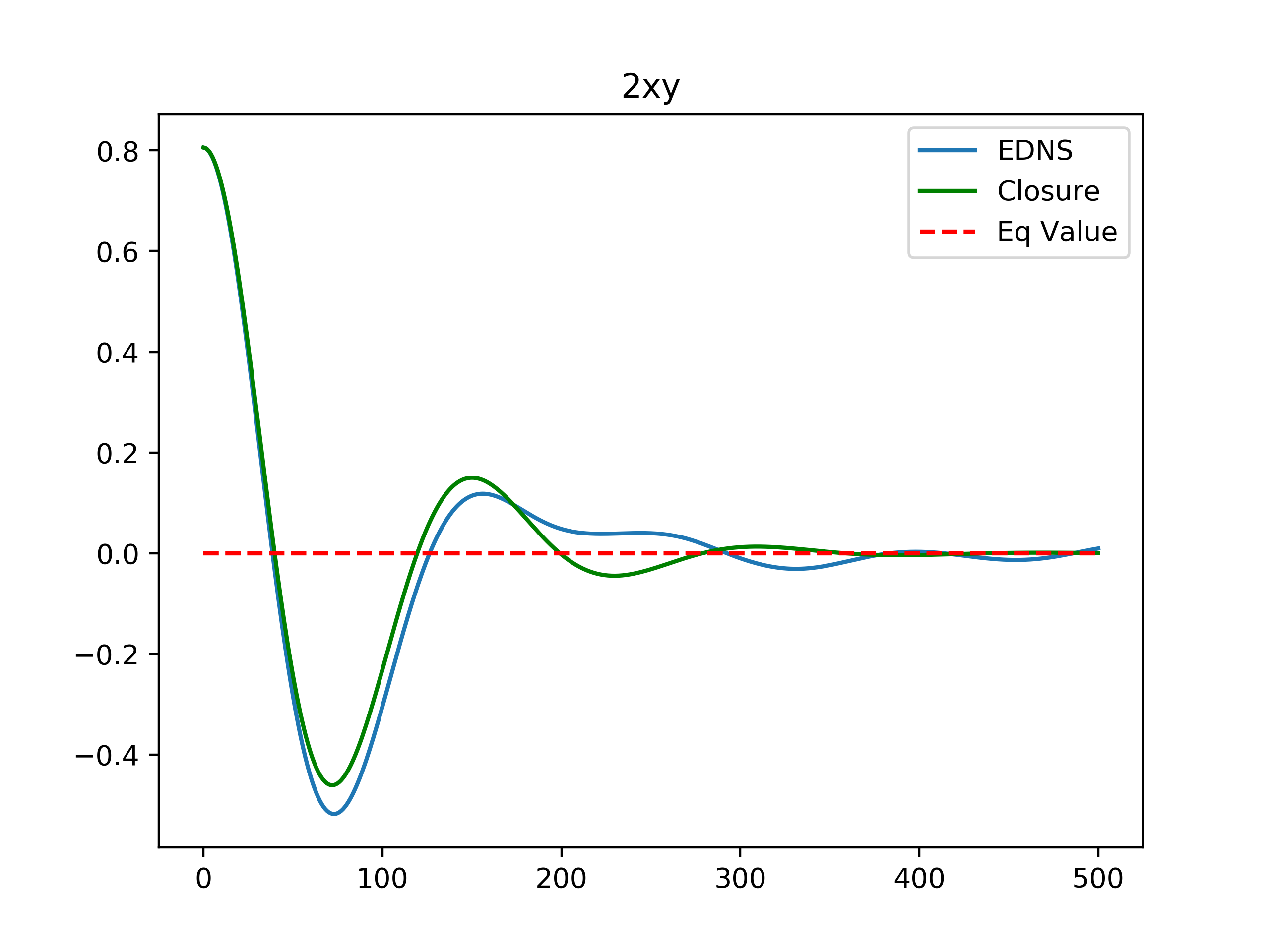}
\end{tabular}
\caption{ Ensemble Direct Numerical Simulation compared to the optimal closure for 
the Euler equations, $R_d=\infty$.  
Left: $a_1(t)=\la x^2-y^2 \ra $; Right:  $a_2(t) = \la 2xy \ra $.  
The initial values are $a_1(0) = 0.0$ and $a_2(0) = 0.8$.}
\label{cov40}
\end{figure}

\begin{figure}[tb]
\begin{tabular}{cc}
	\includegraphics[width=7.2cm]{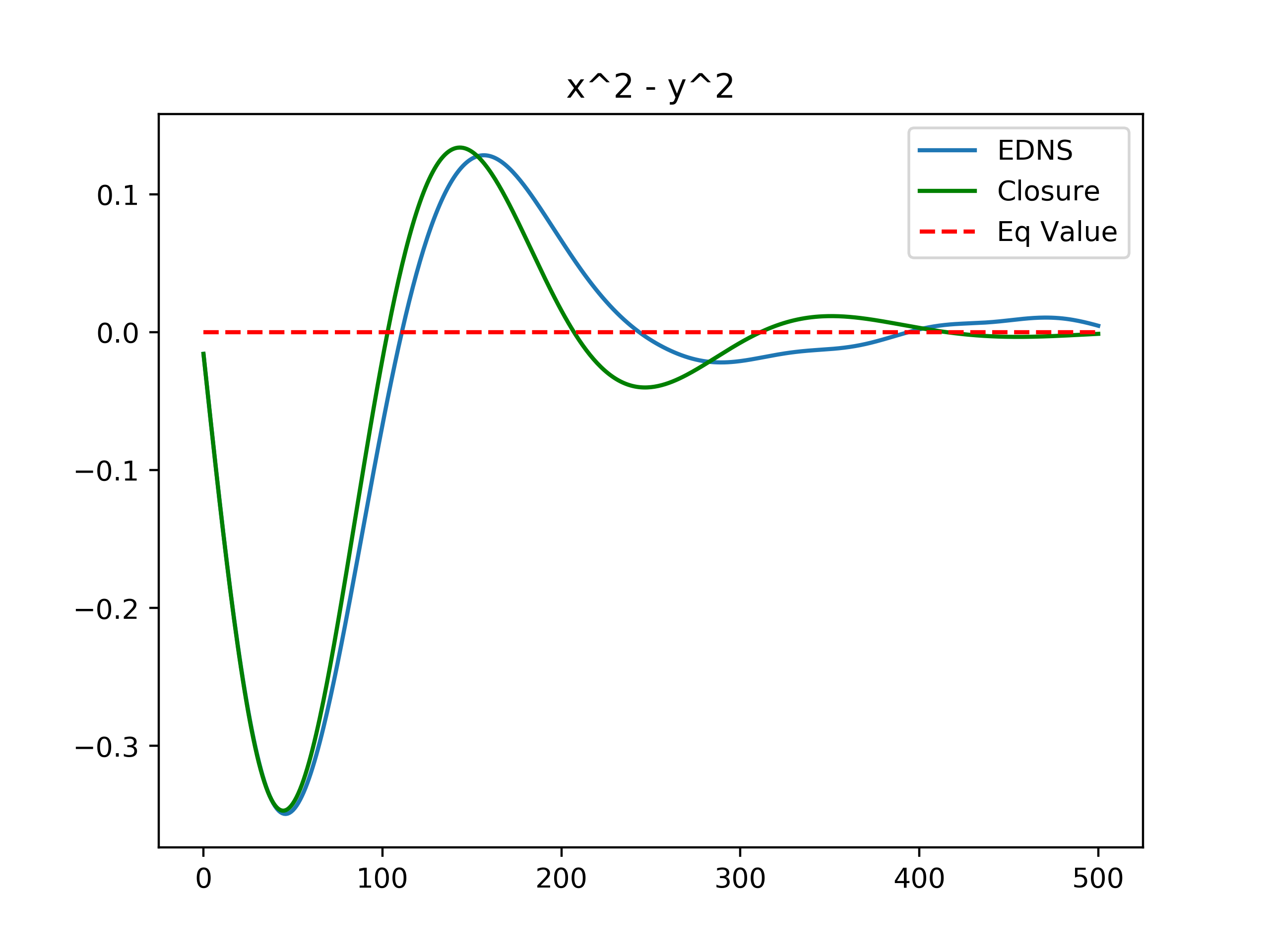} &
	\includegraphics[width=7.2cm]{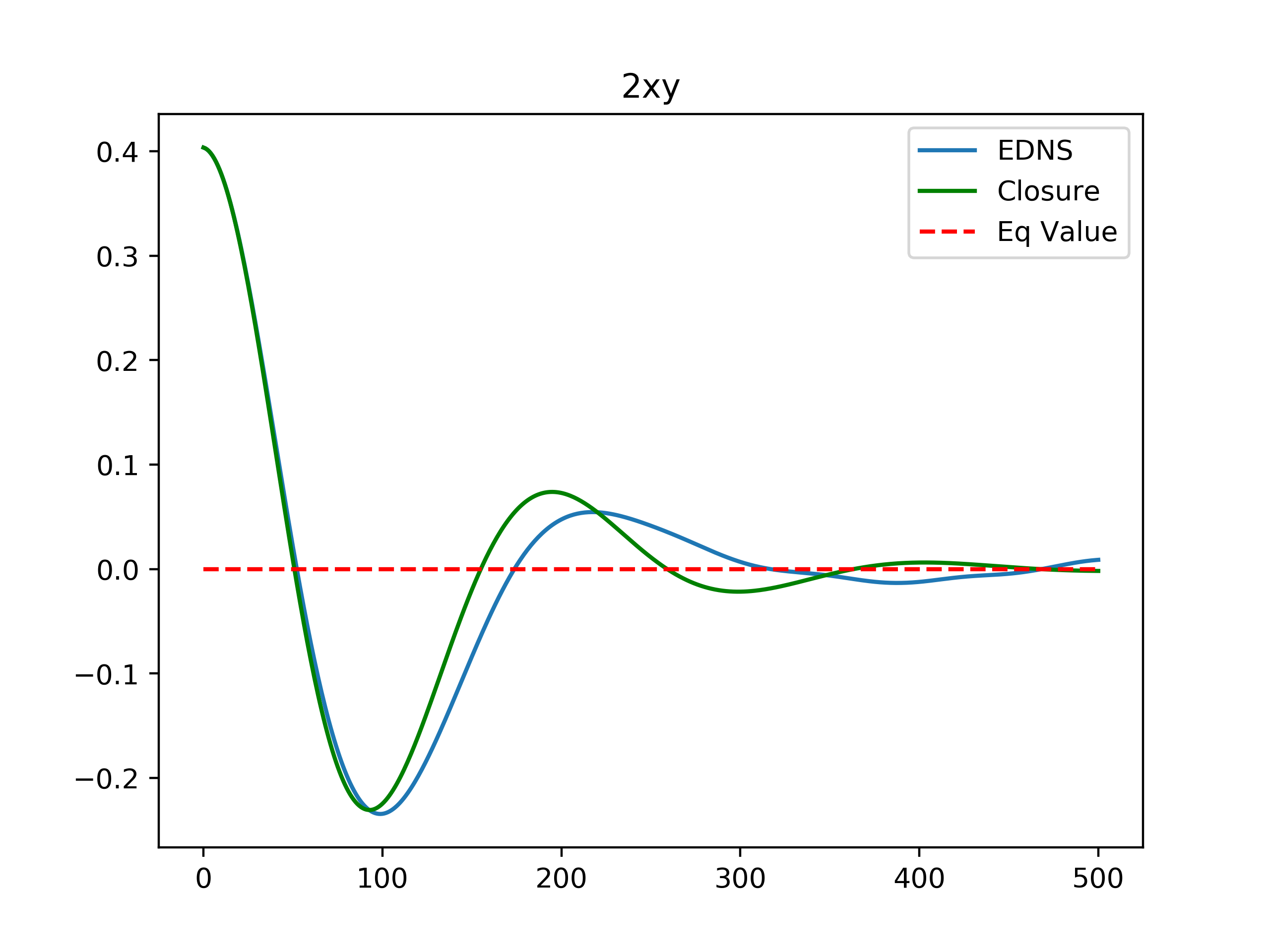}
\end{tabular}
\caption{Ensemble Direct Numerical Simulation compared to optimal closure for single-layer
quasi-geostrophic equations with $R_d = 4.0$.  
Left: $a_1(t)=\la x^2-y^2 \ra $; Right:  $a_2(t) = \la 2xy \ra $.  
 The initial values are $a_1(0)  = 0.0$ and $a_2(0) = 0.4$.}
\label{dr40}
\end{figure}

Figure \ref{cov40} plots the temporal profiles of the two mean resolved variables for  Euler dynamics ($R_d=\infty$).
The initial distortion is relatively large in this case, with $a_1=0, \,  a_2=0.8$.    Nonetheless, a good agreement
with Ensemble Direct Numerical Simulation (EDNS) of 1000 point vortices is exhibited.  In the later stages
of the evolution some departures are apparent, but these can be mainly attributed to finite-$N$ fluctuations.    
Figure \ref{dr40} shows the analogous plots for $R_d= 4$, now with $a_1=0, \, a_2=0.4$.   Comparison to the
full simulation of the vortex ensemble  shows that the optimal closure is less accurate as $R_d$ is decreased.    
Nonetheless, it does capture the time scale of relaxation, which is longer for larger $R_d$.    
This dependence of the rate of relaxation on
$R_d$ is attributed to the effect of screening the vortex-to-vortex interactions.

\begin{figure}[bt]
\centering
\includegraphics[width=10cm]{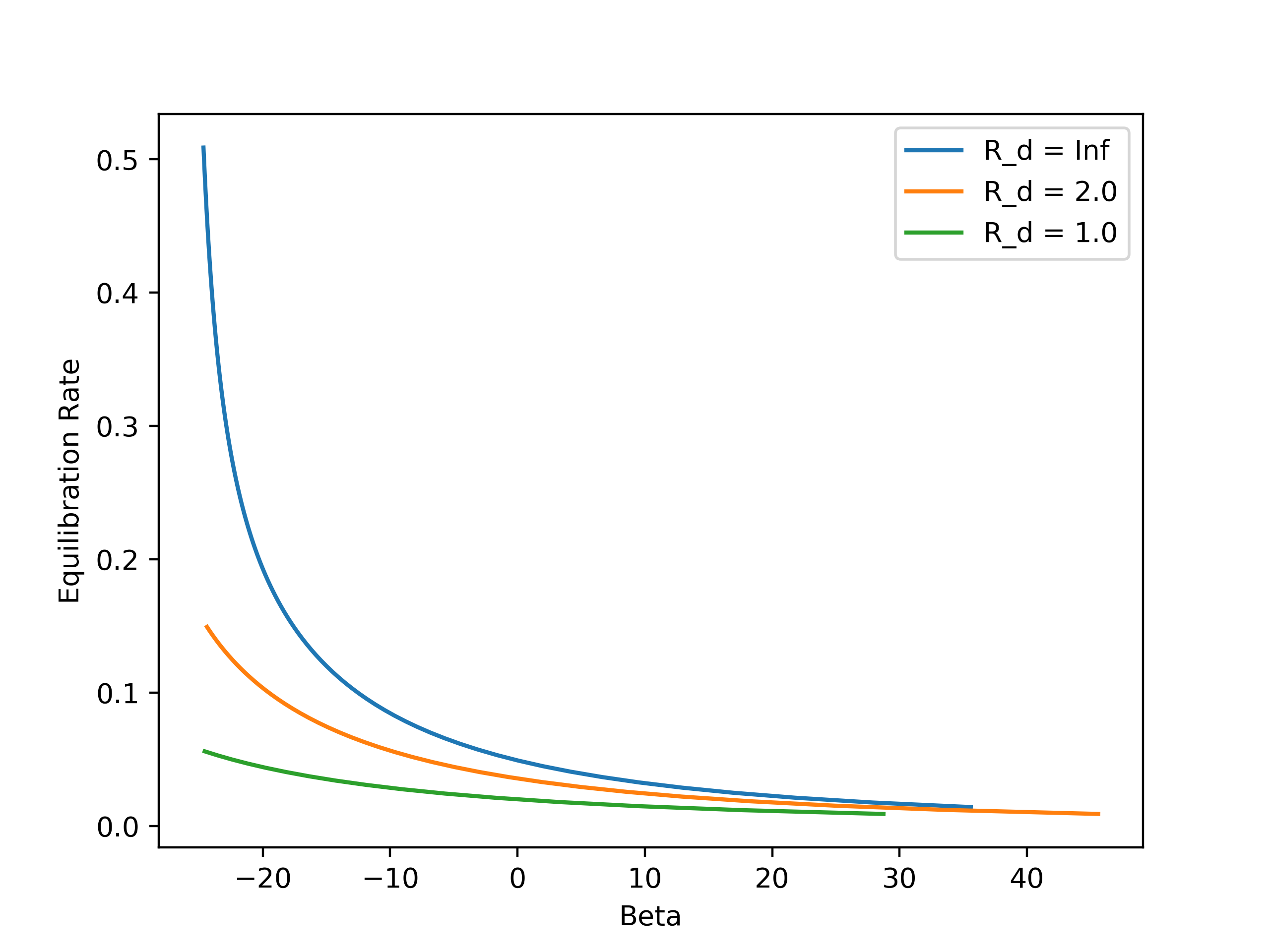} 
\caption{Predicted symmetrization rate versus inverse temperature $\beta$ 
for three values of the deformation radius.  
 As $\beta \rightarrow +\infty $, the equilibration rate tends to zero,
 indicating that symmetrization is suppressed in the deterministic limit.  }
\label{betadissipation}
\end{figure}

\begin{figure}[bt]
\centering
\includegraphics[width=10cm]{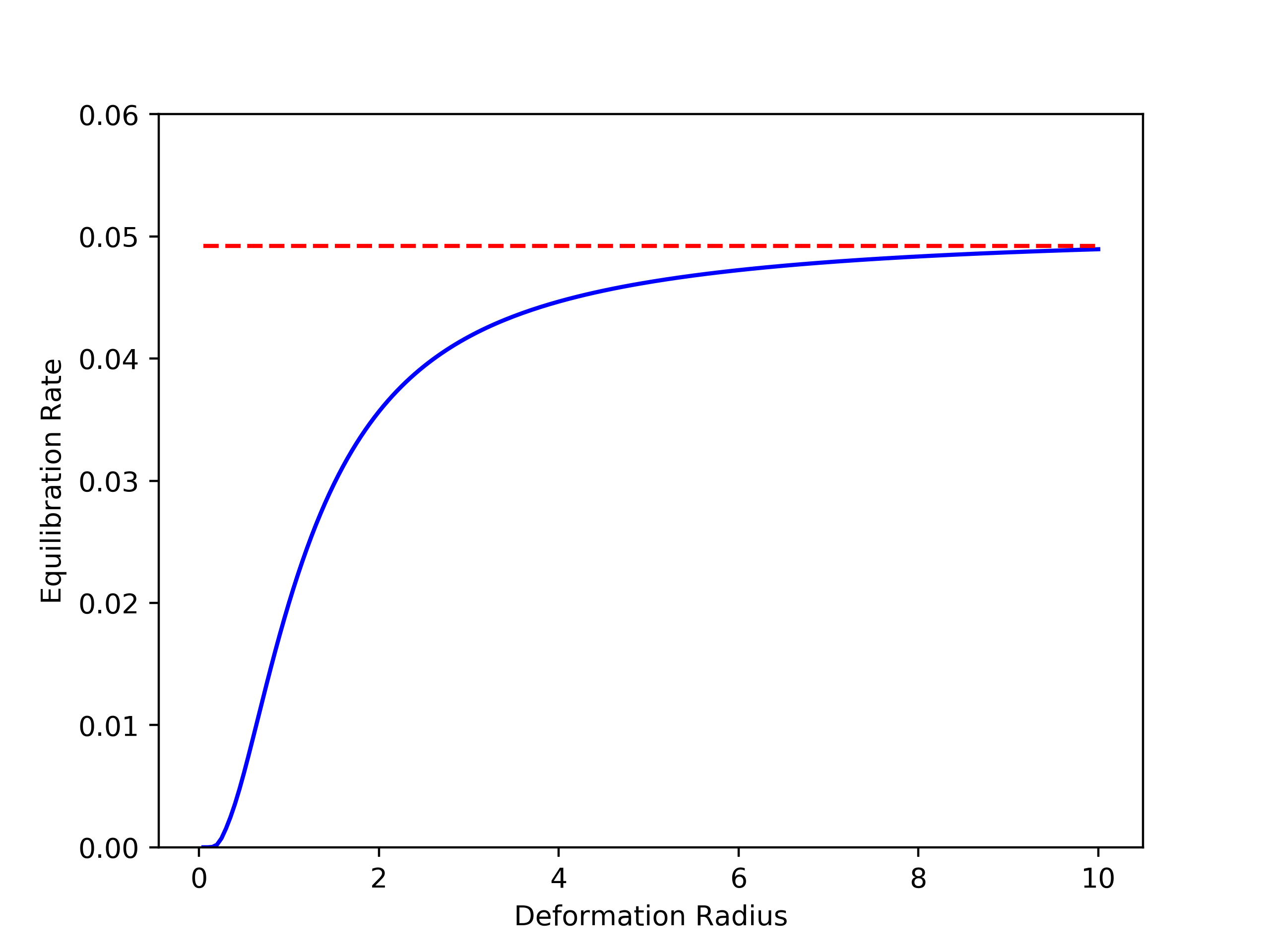}
\caption{Predicted symmetrization rate versus deformation radius for $\beta = 0$.  
The dashed red line is the equilibration value for the Euler equations, $R_d = \infty$.}
\label{rddissipation}
\end{figure}

The dependences of the rate of equilibration on the inverse temperature $\beta_{eq}$ and on the 
Rossby radius $R_d$ are displayed in Figure \ref{betadissipation} and Figure \ref{rddissipation}, respectively. 
The rate plotted in both of these figures is the double eigenvalue of the  $2 \times 2$ matrix $M(+\infty)$,  
the stationary limit of $M(t)$ for large time $t$.    That $M(t)$ is a scalar matrix (diagonal with equal entries)
is a consequence of a rotational symmetry in this reduced model: expressed in polar coordinates
the two resolved variables satisfy,  $A_2(r,\theta) = A_1(r, \theta + \pi/4)$.
This symmetry implies that the coefficient matrices in the Riccati equation (\ref{riccati})
are special:     $C$ and $D$ are scalar, and $J$ is antisymmetric.   (We omit the 
elementary calculations needed to check these properties.)   It follows that the solution
$M(t)$ of (\ref{riccati}) is necessarily a scalar matrix.    

The scalar $M(t)$  has a definite interpretation as the dissipation rate for the reduced equations.   
 To justify this claim we consider the relative entropy $D_{KL}(\zeta_{eq} \, \| \, \zetatilde)$, 
which is the information distance between the nonequilibrium and  equilibrium states.   
Under the near-equilibrium linearization, $D_{KL}(\zeta_{eq} \, \| \, \zetatilde) \approx \xi^T C \xi/2$ .
The closed reduced dynamics (\ref{reduced-eqn-MF}) then yields 
\[
\frac{d}{dt} \frac{1}{2} \xi^T C \xi \, = \, \xi^T \, ( J - M ) \, \xi \, = \,  - \xi^T M \xi \, < 0 \, .   
\]    
Thus the information distance decays at a rate that equals twice the value function, and
hence we naturally identify $M(t) $ as the rate of equilibration.    This rate
is itself time-dependent, since $M(t)$ evolves from $M(0)=0$ to its stationary value $M(+\infty)$,
which solves the associated algebraic Riccati equation.  We therefore choose the 
limiting value of $M(t)$ as $t \rightarrow +\infty$ to define the dissipation rate;  this
stationary value is plotted in Figures \ref{betadissipation}  and \ref{rddissipation}.

 Figure  \ref{betadissipation} shows that the equilibriation rate declines to zero as $\beta_{eq} \rightarrow +\infty$.
 This result has an interesting interpretation.   In this zero-temperature limit, the the equilibrium macrostates
 of the mean-field theory  become deterministic vortex patches (flows having constant potential vorticity 
 inside a closed boundary curve, and zero outside); this limiting property is easily demonstrated from the
 mean-field equation.    For the Euler equations ($R_d=\infty$) 
 an initially elliptical vortex patch does not axisymmetrize, being instead a rotating steady state
  known as a Kirchhoff ellipse (see \S 159 of \cite{lamb}).    Thus,  
 our nonequilibrium mean-field theory is compatible with this classical deterministic result in the zero-temperature limit.    
 The fact that the coherent structures observed in two-dimensional turbulence show a strong tendency to axisymmetrize
 therefore leads us to conclude that typical end states of freely
 decaying turbulence resemble mean-field equilibria having $\beta < 0$,  not large positive $\beta$.

Figure \ref{rddissipation}, in which $\beta_{eq} =0$,  displays the increasing rate of equilibration with 
increasing $R_d$.    This dependence is quite strong when $R_d$ is comparable to the size of the
vortex ensemble, that is, when $R_d \sim 1$.

These results illustrate  the usefulness of a severely coarse-grained model designed to omit the details 
of the reorganizing vorticity field, but to capture key features of its nonequilibrium behavior.   In particular,
the simplicity of this optimal closure allows us to define a unique rate for the axisymmetrization phenomenon,
and to predict its dependence on both the temperature of the equilibrium state and the deformation radius.              

Taking a broader view, the main output of the optimal closure  is 
the matrix $M(t)$, which controls the dissipative structure of the reduced equations.    
The stationary matrix $M(+\infty)$ plays the role of  the transport matrix familiar from  
classical linear irreversible thermodynamics,  in the sense that after the transcient stage it   
 relates the thermodynamic ``forces" $\xi$ to the thermodynamic ``fluxes" $\pi$ \cite{degroot-mazur}.  
From this perspective,
the reduced model tested in this section may be regarded as one instantiation of a nonequilibrium thermodynamics 
of isolated coherent structures.     

\section{Barotropization in Two-Layer  Dynamics}

Finally we apply our optimal closure to a simple model in geophysical fluid dynamics
that includes a vertical stratification.   Specifically, we consider vortex ensembles governed by 
the two-layer quasi-geostrophic equations \cite{pedlosky,salmon}. 
For simplicity, we assume that the two layers have equal depth, and are bounded by a flat bottom and a flat top;
the fluid in the lower layer is denser than the upper layer.       
These shallow layers are separated by a free
surface, and their interaction is controlled by a single parameter, $R_d$, the internal Rossby deformation radius.     
In the appropriate quasi-geostrophic limit, the continuum equations for this system reduce to
 two coupled transport equations for the potential vorticity $q(\bfx,t)$ in each layer;  
the upper layer is indexed by 1, the lower layer by 2:         
\begin{subequations} \label{tlqgsimple}
\begin{align}
	&  \partial_t q_1 + [q_1, \psi_1] = 0,   \;\;\;\;\;\; 
	q_1 = -\Delta \psi_1 +  R_d^{-2} (\psi_1 - \psi_2) \, , 
	 \label{tladvect1} \\
	& \partial_t q_2 + [q_2, \psi_2] = 0,  \;\;\;\;\;\;
	q_2 =   -\Delta \psi_2 - R_d^{-2} (\psi_1 - \psi_2) \, . 
	  \label{tladvect2}
\end{align}
\end{subequations}

Direct numerical simulations of rotating stratified fluids \cite{mcwilliams2} reveal a tendency for coherent vorticity structures to
approach purely barotropic end states, for which  $q_1 \approx q_2$.      This ``barotropization" reflects a preference for minimal potential energy states.  As such it is similar to baroclinic instability, in that it is a mechanism to convert available potential energy into kinetic energy \cite{pedlosky, salmon}.

In light of the physical significance of this tendency toward barotropic states, we now employ our optimal closure theory 
to predict the rate at which initially baroclinic perturbations return to stable barotropic states.  
Our analysis may be viewed as complementary to baroclinic instability theory,
in that we predict the rate of conversion of available potential energy into kinetic energy during the end stages of a 
baroclinic evolution whereas linear stability analysis treats its initial stages.  

We examine only the simplest formulation of this general question.  Namely, we consider two like-signed vortical structures
in the two layers, and demand that they have equal circulation, $\Gamma_1 = \Gamma_2 = 1$.   A barotropic state is one
in which the two vortical structures are vertically aligned.   We therefore create a baroclinic initial state by horizontally displacing the centers of potential vorticity in the two layers.     The mutual advection of these tilted structures as they realign into
a barotropic state may be viewed as a three-dimensional analogue to the axisymmetrization of a single-layer vortical
structure from an elliptically distorted state, examined in the preceding section.     

We take the resolved vector for this problem to be the separation of the centers of the vortex ensembles
in the two layers. In the notation of the previous sections, the ensemble-averaged macrostate $a$ has the components 
$$
	a_1(t)  =  \irr x \, ( \, q_1(\bfx,t) - q_2(\bfx,t) \, ) \, d \bfx \, ,   \;\;\;\;\;\;  
	a_2(t)  =  \irr y \, ( \, q_1(\bfx,t)  - q_2(\bfx,t) \, ) \, d \bfx   \, . 
$$
Thus, the macrostate consists of the first-order spatial moments of the baroclinic part of $q$, while    
the barotropic first-order moments,  $\int \bfx (q_1 + q_2 ) \, d \bfx =0$, are conserved.    
The conserved total energy and angular impulse are, respectively,  
\[
\frac{1}{2} \irr \left( \, \psi_1 q_1 \, + \,  \psi_2 q_2 \,\right) \,  d \bfx  \, = \, E \, , \;\;\;\;\;\;
\irr | \bfx |^2 \, \left( \, q_1   +   q_2 \, \right) \, d \bfx \, = \, L^2 \, .    
\]
Unlike our single-layer reduced models, the second-order moments apart from the $L^2$ are not 
included in the resolved macrostate.    Accordingly, our model is the severest reduction of the two-layer 
dynamics: a mean-field theory that includes only first-order spatial moments of potential vorticity along with
exact invariants.        

The lack-of-fit Lagrangian is derived from a mean-field ansatz that treats the ensembles of vortices 
in each layer as independent.    We omit the straightforward but lengthy calculations needed to formulate 
this optimal closure for the two-layer dynamics, preferring instead to present some predictions
and tests of the ensuing reduced model;  details are available in \cite{maack-thesis}.      

\begin{figure}[bt]
\begin{tabular}{cc}
	\includegraphics[width=7.2cm]{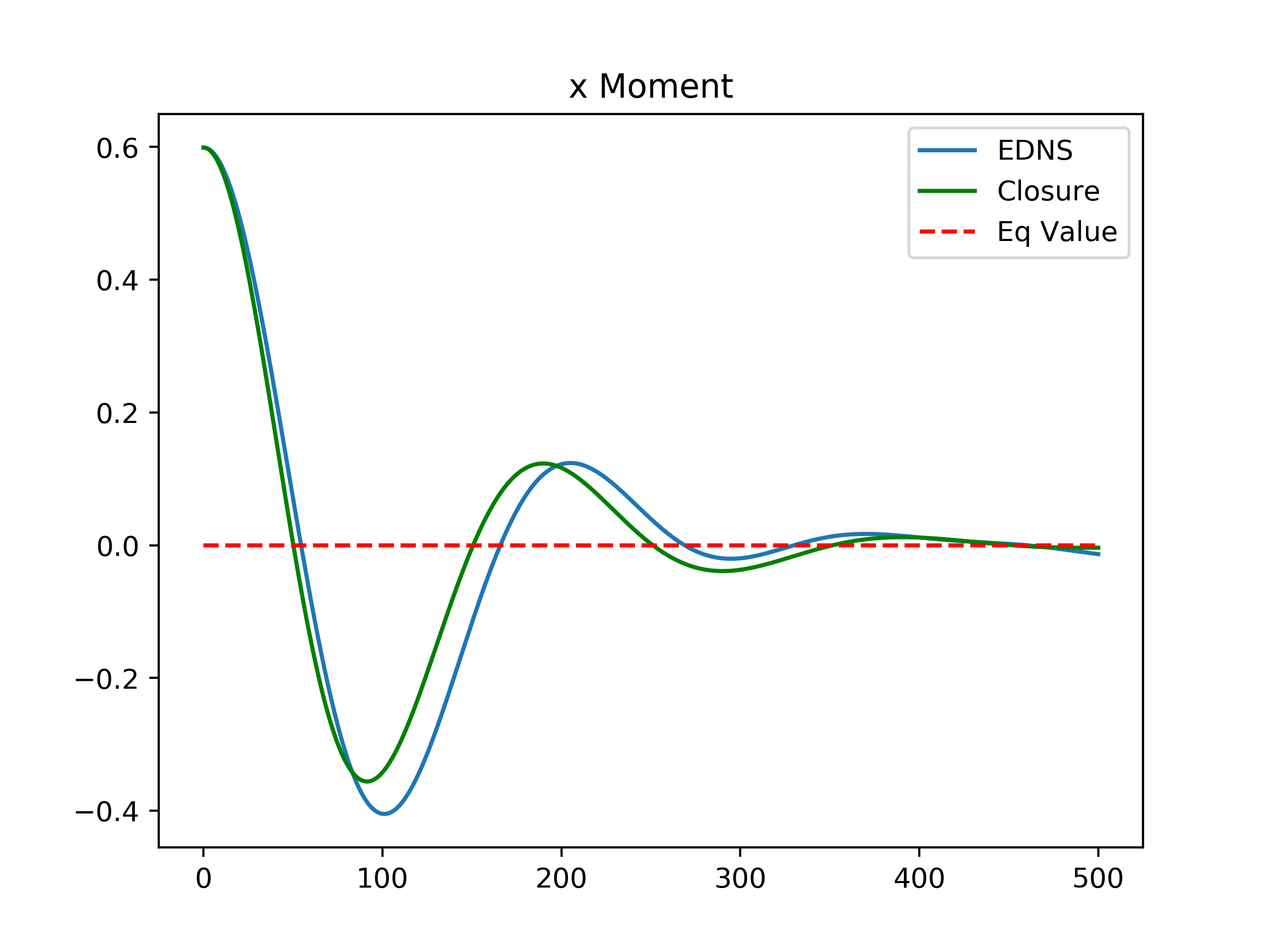} &
	\includegraphics[width=7.2cm]{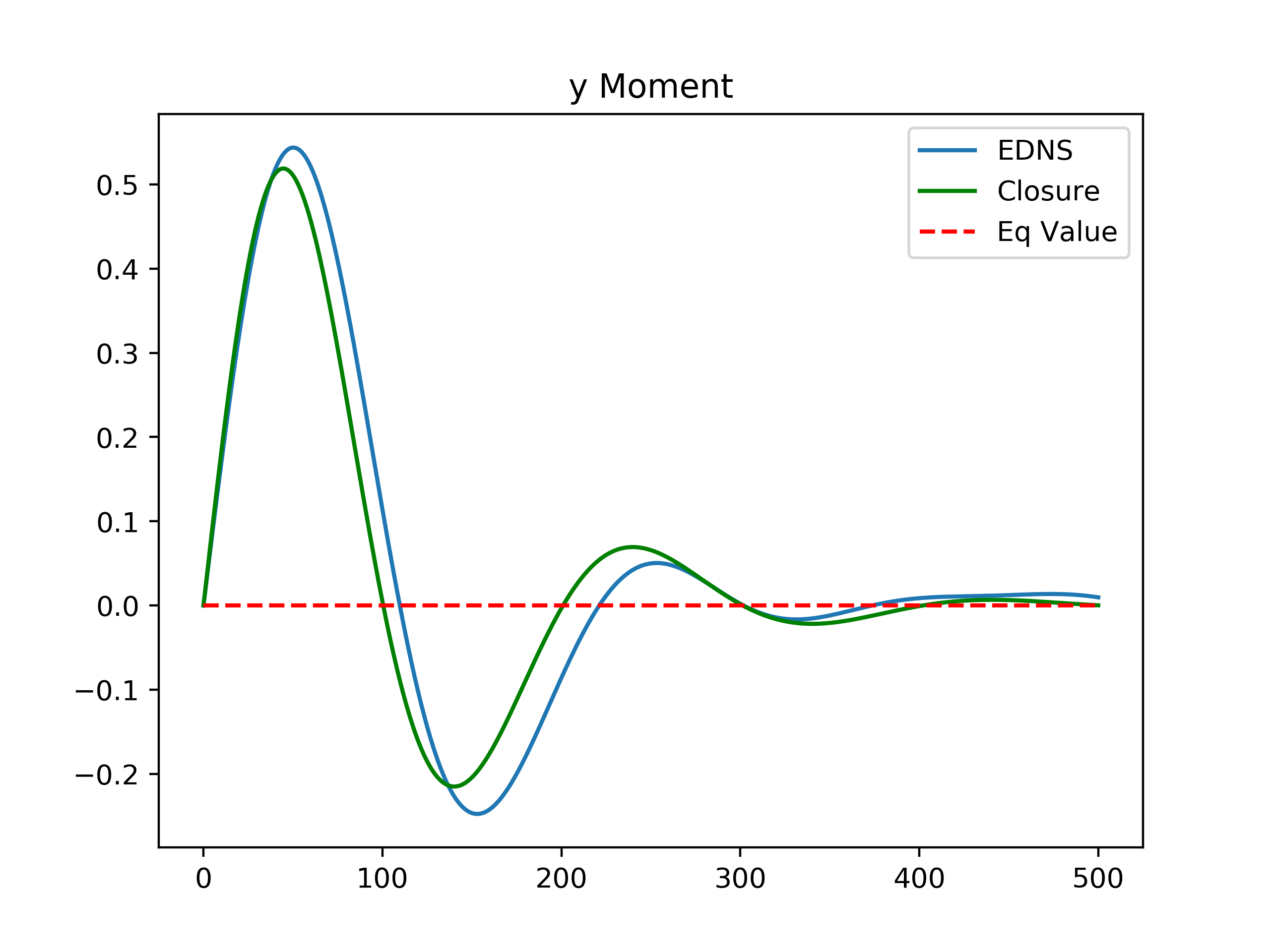}
\end{tabular}
\caption{EDNS compared to optimal closure for two-layer quasi-geostrophic dynamics for  $R_d = 1.0$.
Left: $a_1(t) = \la x_1 - x_2 \ra$;  Right: $a_2(t) = \la y_1 - y_2 \ra$.  
The initial conditions are $a_1 = 0.6$  and $a_2 = 0.0$. }
\label{tilt06}
\end{figure}

Figure \ref{tilt06} compares the optimal closure for initial conditions $a_1 = 0.6, a_2 = 0.0$ with $R_d = 1.0$
against EDNS of a potential vortex system with 1000 vortices in each layer.    The initial state consists of Gaussian
densities in each layer with $L^4 = 4$, chosen so that a Gaussian barotropic state has $x$ and $y$ variances in
each layer normalized to unity.    As the horizontal separation between the center of potential vorticity in each
layer is increased in the initial state,  the  $x$ and $y$ variances are correspondingly decreased 
to maintain the total angular impulse.   Since the size of the vortex ensembles is near to $R_d =1$, there is strong
interaction between the layers and the evolving structure reorganizes into a barotropic equilibrium state relatively
quickly.   Even a moderate increase to $R_d=2$, however, results in a slower equilibration and poorer agreement
between the EDNS and the optimal closure.    The source of this disagreement may be the crudeness of the
closure based only on spatial first moments, or the linearization
of the closure around equilibrium.  It may also be related to memory effects, which are absent from the mean-field
closure used throughout this paper.

\begin{figure}[bt]
\centering
\includegraphics[width=10cm]{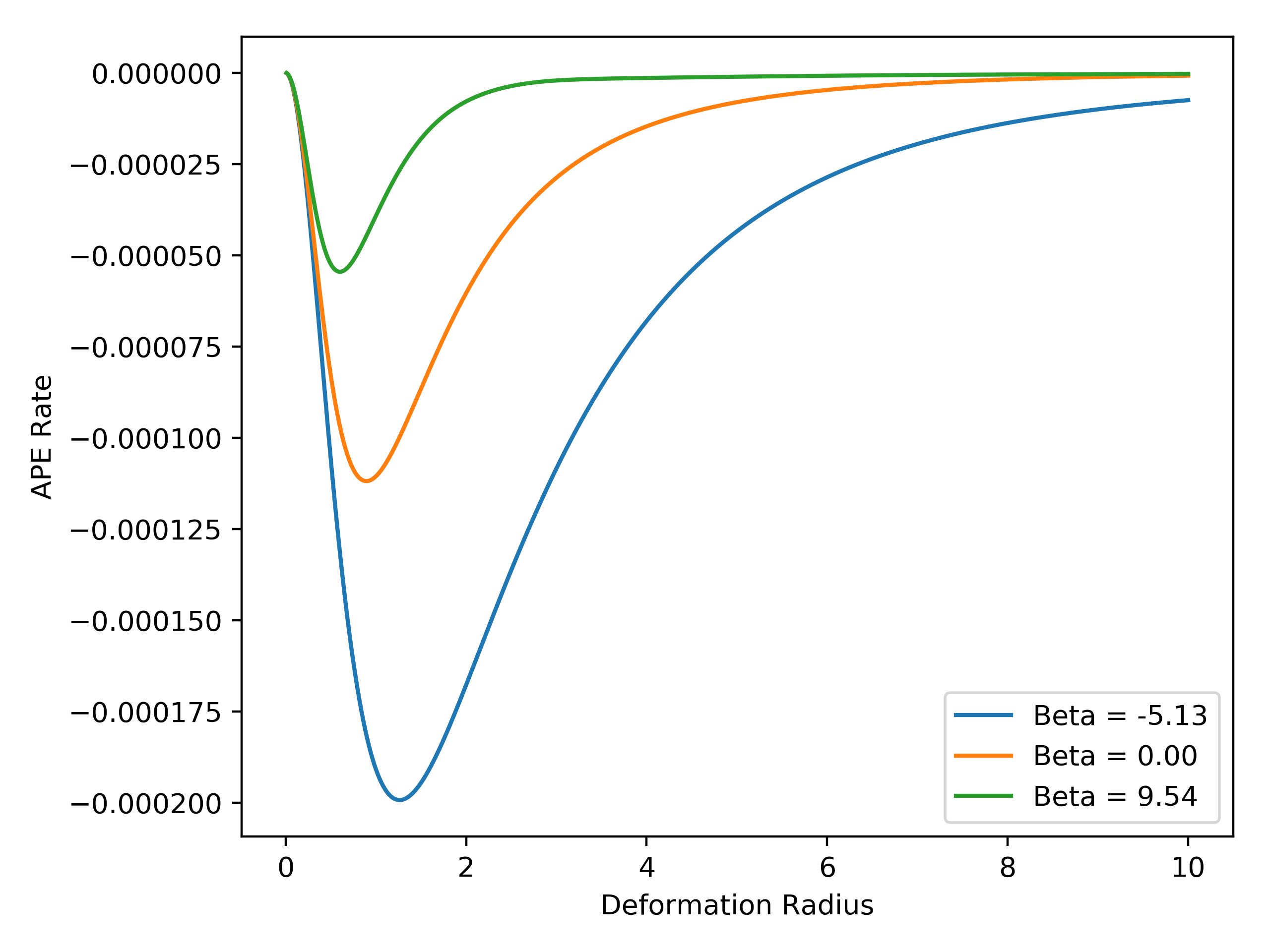}
\caption{Predicted conversion rate of available potential energy as a function of deformation radius $R_d$
 for several values of inverse temperature $\beta$.  
 The peak transfer rates occur when the deformation radius is near to the size of the vortex ensemble.}
\label{tlaperd}
\end{figure}

Figure \ref{tlaperd} displays the dependence of the rate of energy transfer from potential to kinetic energy 
on the inverse temperature $\beta_{eq}$ of the barotropic equilibrium state, and on the internal deformation radius $R_d$.   
The available potential energy is expressible as a quadratic form in $\xi \in \mathbb R^2$, namely,
\[
APE =\frac{1}{2 R_d^2}  \irr ( \psi'_1 - \psi'_2)^2 \, d \bfx \, = \, \frac{1}{2} \xi^T P \xi \, ,     
\]       
where $P$ is the matrix with elements,  
\[
P_{k \ell} = \frac{1}{ R_d^2}  \irr ( \psi'_{1k} - \psi'_{2k}) \, ( \psi'_{1 \ell} - \psi'_{2 \ell}    ) \, d \bfx   \, ;
\]
we recall the notation introduced in (\ref{sensitivities}).     Under the closed reduced dynamics, therefore, 
the rate of conversion of energy is given by 
\[
\frac{d}{dt} APE \, = \, \xi^T P \frac{d \xi}{dt} = \xi^T P C^{-1} (J - M ) \xi \, .    
\]
As in the discussion of the single-layer equilibration rate, the $2 \times 2$ matrices involved in this equation are special, due
to the spatial symmetries of our reduced model;  namely, $C$, $P$ and $M$ are scalar matrices, while $J$ is 
antisymmetric.   The factor $-P C^{-1}M$ therefore controls the rate of conversion of APE, and accordingly 
this scalar factor is displayed in Figure \ref{tlaperd}.  
    
The APE conversion factor is fastest for $\beta<0$;  these barotropic end states are concentrated
coherent structures, and for increasing negative $\beta$ their potential vorticity increasingly concentrates at the origin.    
For $\beta>0$, the APE conversion factor is noticeably smaller, becoming vanishingly small for large positive $\beta$.
For each fixed $\beta$, the fastest conversion occurs when $R_d$ nearly equals the horizontal extent of the
vortex ensemble, which is normalized to near unity in our tests.    This behavior of the reduced model is reminiscent of 
linear stability analysis, which identifies the most active modes as those with length scales close to the deformation
scale  \cite{pedlosky,salmon}.    In this regime our optimal closure successfully
follows the relative displacement of potential vorticity in the two layers and captures the associated 
energy transfer.   Outside of this regime the barotropization process is slower and the reduced model is less accurate.    
Presumably this model is too severely coarse-grained to represent the saturation of a baroclinic perturbation 
accurately over a wide range of deformation scales.  Nonetheless, it is the simplicity of the model 
that allows us define a characteristic rate of energy conversion and thus to identify the conditions 
under which barotropization proceeds most efficiently.     

\section{Conclusion}

The optimal closure method used in this paper to coarse-grain point vortex dynamics is a systematic
approach to model reduction.    From a deterministic microscopic dynamics and a chosen macroscopic description, 
it invokes an information-theoretic metric to characterize that
macroscopic evolution which best fits the underlying microdynamics.    
The metric quantifies the rate of information loss due to statistical reduction 
onto the macroscopic observables.   The closed reduced equations derived from this intrinsic criterion  
rely on no intermediate stochastic modeling or adjustable closure parameters.       
Practical implementation of the optimal closure, however, may be computationally
burdensome, because it requires dynamical optimization over paths of macrostates.      

In the present application we have restricted our attention to relaxation from 
near-equilibrium perturbations, which allows us to linearize the optimal closure around equilibrium states,
thereby making it computationally efficient.    
Motivated by the self-organization of coherent vortex structures observed
in two-dimensional turbulence, we have derived closed reduced equations for the late stage of this organization.
Our closure retains only a few spatial moments of the vorticity, making it a severe coarse-graining
of the vortical structure.
 Comparisons against fully-resolved computations of systems
containing 1000 point vortices have shown that the optimal closure approximates the temporal 
behavior of the few observables retained in the coarse-grained description, and that   
it furnishes a good estimate of the rate of relaxation toward equilibrium.         

The dependence of the relaxation rate on physical parameters is more striking when the fluid dynamics
is extended from the two-dimensional Euler equation to the quasi-geostrophic equations for either the 
single-layer or two-layer shallow water models.   In these geophysical models, the Rossby
radius of deformation, $R_d$ (external or internal depending on the model), is the key length scale.   We have
therefore tested how the reduced models of these quasi-geostrophic vortex systems compare against 
full simulations over a range of deformation scales.    
The axisymmetrization of single-layer structures has been shown to proceed at rates that
decrease with decreasing $R_d$.  The reduced model based on second-order spatial moments approximates
the relaxation toward axisymmetry most accurately for large $R_d$, showing less accuracy as $R_d$ decreases and 
the vortex-vortex interactions weaken.    In the two-layer situation, the reduced
model, which resolves  only the displacement of the centers of vorticity between the two layers,  
predicts the decay of the baroclinic part of the mean flow during relaxation toward a stable barotropic mean flow.    
We have found that this reduced model performs best when the spatial scale of the vortical structure
is near to $R_d$, which is the regime in which the transfer of potential energy 
into kinetic energy is the fastest.  Even though this result pertains to the late stage of relaxation toward a stable structure,
 it harmonizes with the intuition drawn from classical linear 
analysis of baroclinic modes in the early stage of an instability,
in which the fastest growing modes are those whose scale is comparable to $R_d$.       

The reduced models implemented in this paper have used very few resolved variables to
represent the large-scale spatial structures of  macrovortices.   On the one hand,
the severity of this coarse-graining offers a conceptual benefit, because it produces simple nonequilibrium
mean-field equations in which we have been able to identify the relevant rates of relaxation toward
equilibrium.   On the other hand, more refined reduced models would certainly have greater accuracy
over a broader range of physical conditions.     We have therefore 
presented our nonequilibrium mean-field theory of vortex systems for an arbitrary spatial coarse-graining.   
Its range of validity, however, is limited by the mean-field trial densities employed, which rely on an 
independence assumption satisfied by equilibrium states in the appropriate continuum limit, but
not necessarily by general nonequilibrium states.   Moreover, any application of the optimal closure method 
beyond the reach of linearization around equilibrium requires numerical optimization procedures, such as those 
used in optimal control.    The development of optimal closures having wider scope 
than those developed in this paper, therefore, presents challenges for future research.        

\section{Acknowledgments}

This research was funded by the Simons Foundation under grant number 524277.



\begin{thebibliography}{99}

\bibitem{balescu}
R. Balescu,
\newblock {\em Equilibrium and Non-equilibrium Statistical Mechanics,}  
\newblock (Wiley, New York, 1975).   


\bibitem{balian}
R. Balian,
\newblock{\em From Microphysics to Macrophysics: Methods
and Applications of Statistical Physics, I},  
\newblock (Springer-Verlag, 1991).   

\bibitem{boffeta-ecke}
G. Boffetta and R.E. Ecke,
\newblock Two-dimensional turbulence,
\newblock {\it Ann. Rev. Fluid Mech.} 44:427--451 (2012).  

\bibitem{bouchet-sommeria}
F. Bouchet and J. Sommeria,
\newblock Emergence of intense jets and Jupiter's Great Red Spot as maximum-entropy structures.
\newblock {\it J Fluid Mech. } 464: 165--207 (2002)       

\bibitem{bouchet-venaille}
F. Bouchet and A. Venaille,
\newblock Statistical mechanics of two-dimensional and geophysical flows,
\newblock {\it Phys. Rep. } 515: 227--295 (2012).   

\bibitem{CLMP1}
E. Caglioti, P.L. Lions, C. Marchioro and M. Pulvirenti,
A special class of stationary flows for two-dimensional Euler equations: A statistical
mechanics description. I.    
\newblock {\em Commun.  Math. Phys. } 143:501--525 (1992)    

\bibitem{CLMP2}
E. Caglioti, P.L. Lions, C. Marchioro and M. Pulvirenti,
A special class of stationary flows for two-dimensional Euler equations: A statistical
mechanics description. II.    
\newblock {\em Commun.  Math. Phys. } 174: 229--260 (1995)  

\bibitem{casella-berger}
G. Casella and R.L. Berger,
\newblock{ \em Statistical Inference, }
\newblock (Duxbury Press, Belmont, CA, 1990).  


\bibitem{chavanis} 
P.H. Chavanis,  
\newblock Kinetic theory of 2D point vortices from a BBGKY-like hierarchy.   
\newblock {\em Phys. A: Stat. Mech. Appl.}  387(5): 1123--1154 (2008)     

\bibitem{cover-thomas}
T. Cover and J. Thomas,
\newblock{\em Elements of Information Theory,}
\newblock (John Wiley and Sons, New York, 1991).   


\bibitem{degroot-mazur}
S.R. deGroot and P. Mazur,
\newblock {\em Non-equilibrium Thermodynamics,}
\newblock (North Holland, Amsterdam, 1962).   


\bibitem{dibattista-majda}
M.T. DiBattista and A.J. Majda,
\newblock{Equilibrium Statistical Predictions for Baroclinic Vortices: The Role of Angular Momentum}
\newblock{ \it Theoretical and Computational Fluid Dynamics} 14(5): 293--322 (2000)
	
	
\bibitem{ellis-leshouches}
R.S. Ellis,
The theory of large deviations and applications to statistical mechanics.  
\newblock In {\it Long-Range Interacting Systems,} Lecture Notes of the Les Houches Summer School, 90, 
pp. 227--277. T. Dauxois, S. Ruffo and L.F. Cugliandolo, eds. (Oxford U. Press, 2009).  

\bibitem{EHT}
R.S. Ellis, K. Haven and B. Turkington, 
\newblock Nonequivalent statistical equilibrium ensembles and 
refined stability theorems for most probable flows. 
\newblock {\it Nonlinearity} 15:  239-255 (2002).

\bibitem{evans}
L.C. Evans,
\newblock{ \em Partial Differential Equations,}
\newblock  (Amer. Math. Soc., Providence, 1998). 

\bibitem{eyink-spohn}
G. L. Eyink and H. Spohn, Negative temperature states and large-scale
long-lived vortices in two-dimensional turbulence.   
\newblock {\em J. Stat. Phys.} 70: 833--886 (1993)    


\bibitem{gelfand-fomin}
I.M. Gelfand and S.V. Fomin,
\newblock{ \em Calculus of Variations,}
\newblock (Prentice-Hall, Englewood Cliffs, New Jersey, 1963).  

\bibitem{kiessling} M. Kiessling, 
Statistical mechanics of classical particles with logarithmic interactions.   
\newblock {\em Commun. Pure Appl. Math.}  46:27--56 (1993)    

\bibitem{kiessling-lebowitz} M. Kiessling and J. Lebowitz, 
The microcanonical point vortex ensemble: Beyond equivalence.     
\newblock {\em Lett. Math. Phys.}  42: 43--56 (1997)    

\bibitem{kleeman}
R. Kleeman, 
\newblock  A path integral formalism for non-equilibrium Hamiltonian statistical
systems.
\newblock  {\em J. Stat. Phys.} 158: 1271--97 (2015)   


\bibitem{kleeman-turk}
R. Kleeman and B. Turkington,
\newblock A nonequilibrium statistical model of spectrally truncated Burgers-Hopf dynamics.
\newblock  {\em Commun. Pure Appl. Math}  67:1905--46  (2014)   



\bibitem{kraichnan-montgomery}
R.H. Kraichnan and D. Montgomery.
\newblock Two-dimensional turbulence.
\newblock {\em Reports Progr. Phys. } 43: 547--619 (1980) 


\bibitem{kullback}
S. Kullback,
\newblock {\em Information Theory and Statistics,}
\newblock (Wiley, New York, 1959).   

\bibitem{lamb}
H. Lamb,
\newblock{\em Hydrodynamics, $6^{th}$ ed.}
\newblock (Dover Pub., 1945)

\bibitem{liberzon}
D. Liberzon,
\newblock{ \em Calculus of Variations and Optimal Control Theory, A Concise Introduction,} 
\newblock (Princeton U. Press, 2012).  

\bibitem{maack-thesis}
J. Maack,
Reduced models of point vortex systems in quasi-geostrophic fluid dynamics.
\newblock Ph.D.\ dissertation, University of Massachusetts Amherst, May 2018. 

\bibitem{majda-wang}
A.J. Majda and X. Wang, 
\newblock {\it Nonlinear Dynamics and Statistical Theories for Basic Geophysical Flows,}
\newblock (Cambridge U. Press, U.K., 2006).  


\bibitem{marchioro-pulvirenti}
C. Marchioro and M. Pulvirenti, 
\newblock {\em Mathematical Theory of Incompressible Nonviscous Fluids,}
\newblock (Springer-Verlag, 1994).   

\bibitem{mcwilliams1}
J. McWilliams, The emergence of isolated coherent structures in turbulent flow.  
\newblock {\em J. Fluid Mech.} 146:21--43 (1984)

\bibitem{mcwilliams2}
J. McWilliams, Statistical properties of decaying geostrophic turbulence.    
\newblock {\em J. Fluid Mech.} 198: 199--230 (1989)

\bibitem{miller-weichman-cross}
J. Miller, P. Weichman and M.C. Cross,  
Statistical mechanics, Euler's equation, and Jupiters' red spot.   
\newblock {\em Phys. Rev. A } 45: 2328--2359 (1992)   

\bibitem{montgomery-etal}
D. Montgomery, W.H. Matthaeus, D. Martinez and S. Oughton,
Relaxation in two dimensions and the sinh-Poisson equation.  
\newblock {\em Phys. Fluids A} 4: 3--6 (1992)   

\bibitem{newton}
P. Newton, 
\newblock{\em The N-Vortex Problem: Analytical Techniques,} 
\newblock  (Springer, 2001)    

\bibitem{onsager}
L. Onsager, Statistical hydrodynamics.       
\newblock {\em Il Nuovo Cimento} 6: 279--287 (1949)

\bibitem{pedlosky} 
J. Pedlosky,  
\newblock {\em Geophysical Fluid Dynamics,} 
\newblock (Springer-Verlag, 1987)   

\bibitem{robert-sommeria}
R. Robert and J. Sommeria, 
Statistical equilibrium states for two-dimensional flows.  
\newblock {\em J. Fluid Mech.} 229: 291--310 (1991)   

\bibitem{salmon}
R. Salmon,
\newblock {\em Lectures on Geophysical Fluid Dynamics},
\newblock (Oxford Univ. Press, 1998).  

\bibitem{tabeling}
P. Tabeling,
\newblock Two-dimensional turbulence: a physicist approach. 
\newblock {\it Phys. Rep.} 362: 1--62  (2002).    

\bibitem{TT1}
S. Thalabard and B. Turkington,
\newblock Optimal thermalization in a shell model of homogeneous turbulence.  
\newblock {\it J. Phys. A: Math. Theor.} 49(16): 165502 (2016). 

\bibitem{TT2}
S. Thalabard and B. Turkington,
\newblock Optimal response to nonequilibrium disturbances under truncated
Burgers-Hopf dynamics.    
\newblock {\em J. Phys. A: Math. Theor.}  50:  175502 (2017)   


\bibitem{tuckerman}
M.E. Tuckerman, 
\newblock{ \it Statistical Mechanics:  Theory and Molecular Simulation,}
\newblock (Oxford U. Press, U.K., 2010).   

\bibitem{turk-CPAM}
B. Turkington, Statistical equilibrium measures and coherent states in two-dimensional turbulence.  
\newblock  {\em Commun. Pure Appl. Math.}  52:781--809 (1999).      

\bibitem{turk-leshouches}
B. Turkington, 
\newblock Statistical mechanics of two-dimensional and quasi-geostrophic turbulence. 
\newblock In {\it Long-Range Interacting Systems,} Lecture Notes of the Les Houches Summer School, 90, 
pp. 159--209. T. Dauxois, S. Ruffo and L.F. Cugliandolo, eds. (Oxford U. Press, 2009).  

\bibitem{turk-JSP}
B. Turkington, An optimization principle for deriving nonequilibrium statistical
models of Hamiltonian dynamics.  
\newblock  {\em J. Stat. Phys.}  152:569--597 (2013).     

\bibitem{TCT} B. Turkington, Q-Y. Chen and S. Thalabard, 
Coarse-graining two-dimensional turbulence via
dynamical optimization. 
\newblock {\em Nonlinearity}  29:2961-- 2989 (2016)  

\bibitem{TMHD}
B. Turkington,  A. Majda, K. Haven and M. DiBattista, 
Statistical equilibrium predictions of jets and spots on Jupiter.    
\newblock  {\em Proc. Nat. Acad. Sci.}  98:12346--12350 (2001).     

\bibitem{turk-whitaker} B. Turkington and N. Whitaker, 
Statistical equilibrium computations of coherent structures in turbulent shear layers.   
\newblock {\em SIAM J. Sci. Comput.}  17: 1414--1433 (1996)   

\bibitem{wasserman}
L. Wasserman, 
\newblock{\em All of Statistics: A Concise Course in Statistical Inference,}
\newblock (Springer, 2004)   

\bibitem{zwanzig}
R. Zwanzig,
\newblock {\em Nonequilibrium Statistical Mechanics,} 
\newblock (Oxford University Press, New York, 2001). 

\end{thebibliography}



\end{document}